\documentclass[aps,pre,reprint,amsmath,amssymb,ap,superscriptaddress]{revtex4-2}
\usepackage[utf8]{inputenc}
\usepackage[T1]{fontenc}
\usepackage{graphicx}
\usepackage{float}
\usepackage{color}
\usepackage{xr}
\usepackage{bm}
\usepackage{rotating}
\usepackage{ulem}
\usepackage{hyperref}

\begin{document}

\preprint{APS/123-QED}

\title{Hyperuniformity in ternary fluid mixtures: the role of wetting and hydrodynamics}

\author{Nadia Bihari Padhan}
 \email[]{nadia\_bihari.padhan@tu-dresden.de}
\affiliation{
 Institute of Scientific Computing, TU Dresden, 01069 Dresden, Germany
}
\author{Axel Voigt}
 \email[]{axel.voigt@tu-dresden.de}
\affiliation{
 Institute of Scientific Computing, TU Dresden, 01069 Dresden, Germany
}
\affiliation{
Center of Systems Biology Dresden (CSBD), Pfotenhauerstr. 108, 01307 Dresden, Germany
}
\affiliation{
 Cluster of Excellence, Physics of Life (PoL), 
 TU Dresden, Arnoldstr. 18, 01307 Dresden, Germany
}

\date{\today}

\begin{abstract}
Phase separation in multicomponent fluids is central to understanding the organization of complex materials and biological structures. The Cahn-Hilliard-Navier-Stokes (CHNS) equations offer a robust framework for modeling such systems, capturing both diffusive dynamics and hydrodynamic interactions. In this work, we investigate hyperuniformity—characterized by suppressed large-scale density fluctuations—in ternary fluid mixtures governed by the ternary CHNS equations. Using large-scale direct numerical simulations, we systematically explore the influence of wetting conditions and hydrodynamic effects on emergent hyperuniformity. Similar to binary systems we observe that the presence of hydrodynamics weakens the hyperuniform characteristics. However, also the wetting properties have an effect. We find that in partial wetting regimes, all three components exhibit comparable degrees of hyperuniformity. In contrast, for complete wetting scenarios, where one component preferentially wets the other two, the wetting component displays a significant reduction in hyperuniformity relative to the others. These findings suggest that wetting asymmetry can act as a control parameter for spatial order in multicomponent fluids.
\end{abstract}

\maketitle
\section{Introduction}
Phase separation is ubiquitous in nature and emerges as a universal phenomenon across a variety of physical settings---ranging from multiphase fluid mixtures~\cite{zarzar2015dynamically,shono2023spontaneous,yanagisawa2007growth, cai2021fluid, iqbal2013controlling,clegg2016one} and active matter systems~\cite{Cates_2025, bhattacharyya2023phase, zhang2021active, zheng2024universal,cates2010arrested}, to liquid–liquid phase separation in biology~\cite{gouveia2022capillary,wilken2023spatial,bansal2022active,saha2022active,kaur2021sequence,hyman2014liquid,balasubramaniam2021investigating}. In phase separation, an initially homogeneous mixture becomes unstable and evolves into domains of nearly pure phases~\cite{bray1994theory,chaikin1995principles}. This process gives rise to diverse morphologies: from spinodal bicontinuous structures in binary mixtures~\cite{cates2018theories}, to microphase-separated patterns in active matter~\cite{cates2015motility,Cates_2025}, to complex emulsion structures in ternary fluid mixtures~\cite{zarzar2015dynamically,mao2020designing,shek2022spontaneous} and ternary biomolecular systems~\cite{kaur2021sequence,gouveia2022capillary}. The dynamics of these morphologies is typically governed by the properties of the interfaces between the phases.  While a binary mixture has only one interface, a ternary mixture involves three distinct interfaces, where asymmetries in interfacial tensions and wetting preferences introduce additional complexity and lead to a richer variety of morphologies. This increase in complexity increases even further if the number of components grows. 

We first review results for binary mixtures. They are relatively well understood and their resulting morphologies exhibit characteristics of hyperuniformity~\cite{ma2017random}---a hidden form of spatial order in a disordered material, which is characterized by suppressed density fluctuations at large scales~\cite{Torquato2003}. Such hyperuniform systems somehow combine the long-range order of crystals with the short-range isotropy of liquids. They are appealing, as they are resilient against imperfections, defects, and, to some extent, noise, which are peculiar properties in various applications, see~\cite{Tor18} for a comprehensive review. Hyperuniformity is consistently reproduced in passive and active scalar-field theories based on the Cahn–Hilliard (CH) equations~\cite{de2024hyperuniformity,zheng2024universal,ma2017random,padhan2025suppression} and also experimentally hyperuniformity in binary systems has been confirmed in patterns of DNA droplets~\cite{wilken2023spatial}. Given that ternary phase separating systems feature more complex interfacial dynamics than their binary counterparts, it is natural to ask whether hyperuniformity persists in these systems and how material properties influence it. Surprisingly, these questions remain largely unexplored, even if ternary phase separation is gaining increasing attention.

In this work, we investigate these fundamental issues using direct numerical simulations of ternary phase-field models, the ternary Cahn-Hilliard (CH) equations and the ternary Cahn-Hilliard-Navier-Stokes (CHNS) equations. We systematically examine how wetting conditions and hydrodynamic interactions impact the emergence of hyperuniformity during phase separation in ternary mixtures. Our results reveal that hydrodynamics tend to suppress hyperuniformity, an effect that becomes more pronounced with the inclusion of inertia, where large-scale fluctuations are enhanced. The system displays a rich variety of morphologies, including interconnected droplets and nested double emulsions—structures reminiscent of biological phase separation and immiscible ternary fluid systems. We find that in partial wetting regimes, all three components display comparable degrees of hyperuniformity. In contrast, under complete wetting conditions---where one component preferentially wets the other two---the wetting component shows a marked reduction in hyperuniformity. 

The paper is structured as follows: In Sec. \ref{sec…2} we introduce the considered model and describe the used numerical approach to solve it. In Sec. \ref{sec…3} we discuss the results and in Sec. \ref{sec:4} we draw conclusions. 

\section{Model and Numerical methods} \label{sec…2}

The mathematical modeling of fluid mixtures with more than two components is significantly more complex than their binary counterparts. In such systems, one must account for multiple interfaces, each characterized by distinct surface tensions, and accommodate differences in fluid properties such as viscosity and density among the components. Over the past decade, phase-field-based continuum models have emerged as a powerful framework for describing $n$-component immiscible fluid mixtures~\cite{boyer2014hierarchy,dong2014efficient,nurnberg2017numerical,dong2018multiphase,dong2017wall,mao2019phase,mao2020designing,abels2024mathematical}. Among them, particular attention has been given to incompressible ternary fluid mixtures, which capture essential features of multiphase systems with three interacting components~\cite{kim2005phase,kim2012phase,boyer2010cahn,boyer2006study,hester2023fluid,liang2016lattice,abadi2018conservative,yang2021new,kim2007phase,semprebon2016ternary,padhan2025cahn,wohrwag2018ternary,park2012ternary,shen2023bubble,park2016diffuse}. In this work, we adopt the model proposed by Boyer et al.~\cite{boyer2006study}, along with the assumption of equal and constant density for all components. Under this assumption, the model becomes equivalent to a special case of the more general hydrodynamic framework developed by Dong et al.~\cite{dong2018multiphase}, which also provides the basis for the mathematically analysis of the model considered in \cite{abels2024mathematical}.

\subsection{Ternary Cahn-Hilliard-Navier-Stokes framework}
\begin{figure}
{
\includegraphics[width=\linewidth]{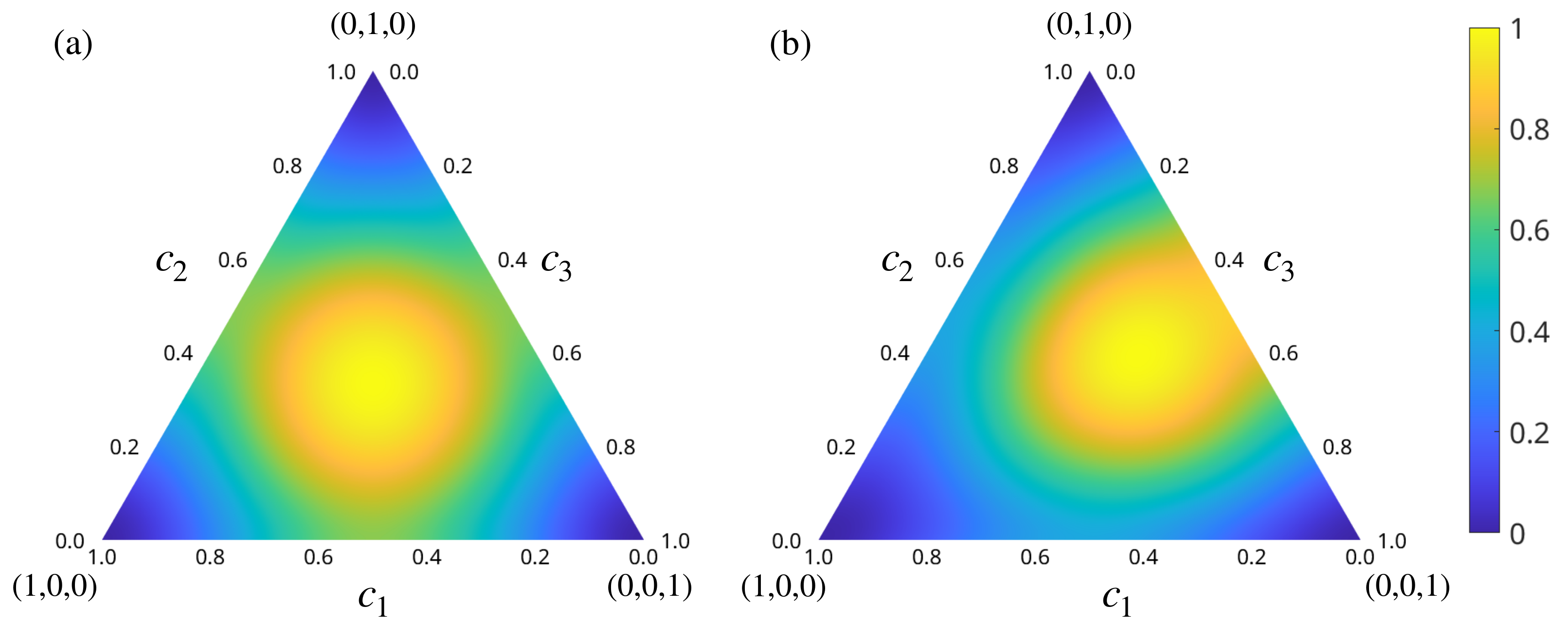}
}
\caption{\label{fig:fr_en} 
{Pseudocolor plots of free energy landscape [eq.~\ref{eq:free-energy}], projected onto a composition triangle, for (a) partial wetting with surface tensions $(\sigma_{12}, \sigma_{23}, \sigma_{13}) = (0.45, 0.45, 0.45)$, and (b) complete wetting with $(\sigma_{12}, \sigma_{23}, \sigma_{13}) = (0.2, 0.45, 0.2)$, where phase-2 preferentially wets phases $1$ and $3$. The three vertices of the triangle correspond to the pure phases $(c_1,c_2,c_3) = \{(1, 0, 0), (0,1,0), (0,0,1)\}$, and the three edges represent the composition axes. The colorbar indicates the free energy $F(\{c\})$, normalized by its maximum value. The positive constant $\Lambda$ is fixed at $6$.
}
}
\end{figure}

We consider an incompressible, immiscible three-component fluid mixture, where each component is described by a scalar concentration field $c_i(\mathbf{x})$, subject to the constraint $\sum_{i=1}^{3} c_i(\mathbf{x}) = 1$. For such complex multicomponent systems, we define the following Landau-Ginzburg-type free-energy functional $\displaystyle \mathcal F (\{c_i, \nabla c_i\})$ for three-component fluid mixture~\cite{boyer2006study}:
\begin{eqnarray}
\hspace{-0.35cm}
\mathcal{F}(\{c_i, \nabla c_i\}) = \int\displaylimits_{\Omega} d\Omega \Bigg[
\frac{12}{\epsilon} F(\{c\}) 
+ \frac{3\epsilon}{8} \sum_{i=1}^{3} \gamma_i (\nabla c_i)^2 \Bigg].
\label{eq:functional}
\end{eqnarray}
Since the system consists of three immiscible fluids, there are three distinct interfaces between the fluid components. The interfacial tension coefficient for the $(ij)_{th}$ interface is defined as 
\begin{eqnarray}
    \sigma_{ij} = (\gamma_i + \gamma_j) / 2\,.\label{eq:st}
\end{eqnarray}
We assume all interfaces to have the same width $\epsilon$. The Helmholtz free-energy $F(\{c\})$ with three minima at $(c_1, c_2, c_3) = (1, 0, 0), (0,1,0), (0,0,1)$---corresponding to the three pure phases---is given by 
\begin{eqnarray}
F(\{c\}) = \sum_{i=1}^{3} \frac{\gamma_i}{2} c_i^2 (1-c_i)^2 + \Lambda c_1^2 c_2^2 c_3^2\,.
\label{eq:free-energy}
\end{eqnarray}
Following Refs.~\cite{boyer2006study,boyer2010cahn}, we include a higher-order term in the free energy, with coefficient $\Lambda > 0$, to ensure thermodynamic consistency and numerical stability by keeping the free energy bounded from below. This term becomes particularly useful when one of the coefficients $\gamma_i$ is negative, which can occur under complete wetting conditions where one surface tension coefficient is significantly larger than the other two. Note that $\sigma_{ij} > 0$ always holds, individual $\gamma_i$ can be negative to promote selective wetting. In Fig.~\ref{fig:fr_en}, we illustrate the pseudocolor plots of free energy, eq.~\eqref{eq:free-energy}, for two sets of surface tensions coefficients: (a) $(\sigma_{12}, \sigma_{23}, \sigma_{13}) = (0.45, 0.45, 0.45)$ and (b) $(\sigma_{12}, \sigma_{23}, \sigma_{13}) = (0.2, 0.45, 0.2)$, where the former correspond to partial wetting and the latter to complete wetting. In the complete wetting case, the free energy landscape indicates a high energetic cost associated with forming 
$23$-interfaces, leading the system to favor phase $1$ as an intermediate layer that wets both phases $2$ and $3$—a behaviour we observe and elaborate on in the results section. The type of wetting can be further formalized by defining the wetting parameter~\cite{quere1990spreading,leermakers2025no,kovalchuk2024spreading}
\begin{eqnarray}
    W_i = \sigma_{jk} - (\sigma_{ij} + \sigma_{ik}), \, i \neq j \neq k\,,\label{eq:wetting}
\end{eqnarray}
where the phase-$i$ is said to wet phases $j$ and $k$ when $W_i > 0$. The ternary mixture is governed by the following coupled (incompressible) Cahn–Hilliard–Navier–Stokes (CHNS) equations:
\begin{eqnarray}
    \partial_t \mathbf{u} + ({\mathbf u}\cdot {\nabla}){{ \mathbf u}} &=& \nu {\nabla}^2
{{\mathbf u}} - \nabla P + \mathbf{F^{int}}\label{eq:CHNS-A} \\
\nabla \cdot \mathbf{u} &=& 0\label{eq:incompress}\\
 \partial_t{c_{i}} + (\mathbf{u}.{\nabla})c_{i} &=& \frac{M}{\gamma_{i}}{\nabla}^2 \mu_{i}, \;\; i = 1, 2 .
\label{eq:CHNS-B}
\end{eqnarray}
Thereby $\mathbf{u}(\mathbf{x},t)$ denotes the velocity and $P(\mathbf{x},t)$ the pressure, $\nu$ is the (constant) kinematic viscosity, $M$ a (constant) mobility and
\begin{eqnarray}
   \mathbf{F^{int}} =  {\sum_{i=1}^{3}\mu_{i} {\nabla} c_{i}}
   \label{eq:force}
\end{eqnarray}
the interfacial force density. We evolve only two concentration fields, $c_1$ and $c_2$, since $c_3$ can be recovered from the relation $\sum_{i=1}^{3} c_i = 1$. For our study, we assume that the mixture has a constant density, set to unity, and all three components possess equal kinematic viscosity $\nu$. This choice is sufficient for our purposes and does not contradict physical consistency, as we focus exclusively on phenomena governed by interfacial dynamics. The chemical potential is defined as the variational derivative of the functional (Eq.~\ref{eq:functional}) as $\mu_i = \frac{\delta \mathcal F}{\delta c_i} + \beta(\{c\})$, where $\beta(\{c\})$ is a Lagrange multiplier used to satisfy the constraint $\sum_{i=1}^{3} c_i = 1$; the form of $\beta(\{c\})$ can be obtained from Eq.~\ref{eq:CHNS-B}. This leads to the following form of the chemical potential for the $i_{th}$ component:
\begin{eqnarray}
    \mu_i &=& G(\{c\}) - \frac{3}{4} \epsilon \gamma_i \nabla^2 c_i \\
    G(\{c\}) &=& \frac{4 \gamma_{\scriptscriptstyle \mathrm{T}}}{\epsilon} \sum_{j\neq i} \frac{1}{\gamma_j} \Bigg(\frac{\partial F(\{c\})}{\partial c_i}- \frac{\partial F(\{c\})}{\partial c_j}\Bigg),
\end{eqnarray}
with $3/\gamma_{\scriptscriptstyle \mathrm{T}} = \sum_{i=1}^{3} 1/\gamma_i$. All components of $\mu_i$ are linked through the relation $\sum_{i=1}^{3} \mu_i/\gamma_i = 0$~\cite{boyer2006study}.

\subsection{Numerical methods}
We write Eqs. (\ref{eq:CHNS-A})-(\ref{eq:CHNS-B}) in the following vorticity-streamfunction formulation, where $\omega = \mathbf{e}_3 \cdot \bm{\omega}$ is the out-of-plane vorticity, and it is related to the streamfunction $\Phi$ via the velocity $\mathbf{u}$ as $\omega = (\nabla \times \mathbf{u})\cdot \mathbf{e}_3$ and $\mathbf{u} = \nabla \times (\Phi \mathbf{e}_3)$. We obtain: 
\begin{eqnarray}
\partial_t{{\omega}} + ({\mathbf u}\cdot {\nabla}){{ \omega}} &=& \nu {\nabla}^2
{{\omega}} + [{\nabla} \times \mathbf{F^{int}}]\cdot {\mathbf{e}}_3 \label{eq:strfn1}\\
\partial_t{c_{i}} + (\mathbf{u}.{\nabla})c_{i} &=& \frac{M}{\gamma_{i}}{\nabla}^2 \mu_{i}, \;\; i = 1, 2 \label{eq:strfn2} \\
\mu_i &=& G(\{c\}) - \frac{3}{4} \epsilon \gamma_i \nabla^2 c_i , \;\; i = 1, 2 \,.
\label{eq:mu}
\end{eqnarray}
This formulation naturally satisfies the incompressible condition $\nabla \cdot \mathbf{u} = 0$ and eliminates the pressure $P$. We perform Fourier pseudospectral-based direct numerical simulations~\cite{padhan2025cahn} of Eqs.~(\ref{eq:strfn1})-(\ref{eq:mu}) in a two-dimensional periodic box with side length $4\pi$ and $1024^2$ collocation points. First we write Eqs.~(\ref{eq:strfn1})-(\ref{eq:strfn2}) in Fourier space using the Fourier coefficients $\hat{c}_i(\mathbf{k}, t)$, $\hat{\omega}(\mathbf{k}, t)$, $\hat{\Phi}(\mathbf{k}, t)$, $\hat{\mathbf{u}}(\mathbf{k}, t)$ as follows:
\begin{eqnarray}
\partial_t\hat{\omega}(\mathbf{k}, t) &=& -\widehat{(\mathbf{u}\cdot \nabla \omega)}(\mathbf{k}, t) - \nu |\mathbf{k}|^2 \hat{\omega}(\mathbf{k}, t) \label{eq:NS-A}\\
&& + \dot{\iota} \mathbf{k} \times \hat{\mathbf{F}}^{\mathbf{int}}(\mathbf{k}, t) \nonumber\\
\hat{\Phi}(\mathbf{k}, t) &=& \hat{\omega}(\mathbf{k}, t)/|\mathbf{k}|^2\,;\;\; \hat{\mathbf{u}}(\mathbf{k}, t) = \dot{\iota} \mathbf{k} \times \hat{\Phi}(\mathbf{k}, t) \label{eq:NS-B}\\
\partial_t \hat{c}_i(\mathbf{k},t) &=& -(\widehat{\mathbf{u}\cdot \nabla c_i})(\mathbf{k}, t) -\frac{3}{4} M \epsilon |\mathbf{k}|^4 \hat{c}_i(\mathbf{k}, t) \label{eq:CH}\\&& - \frac{M}{\gamma_i} |\mathbf{k}|^2 \widehat{G(\{c\})}(\mathbf{k}, t)\nonumber\,.
\end{eqnarray}
To circumvent the computational cost of convolutions, we compute all nonlinear terms in Eqs.~(\ref{eq:NS-A}) and (\ref{eq:CH}) in real space and subsequently transform them to Fourier space. We apply a standard $1/2$-dealiasing scheme to suppress aliasing errors introduced by the nonlinear terms. We integrate Eqs.~(\ref{eq:NS-A}) and (\ref{eq:CH}) using the exponential time differencing Runge-Kutta method of second order (ETDRK2)~\cite{cox2002exponential}, where the linear terms in $\hat{\omega}(\mathbf{k})$ and $\hat{c}_i(\mathbf{k})$ are treated exactly via their exponential solutions, while the nonlinear terms are evaluated in a semi-implicit manner using linear integrating factors. We leverage high-performance computing using a Message Passing Interface (MPI) framework in our custom-developed C code, which employs the FFTW library for optimized Fast Fourier Transform (FFT) operations. We consider the following initial conditions for $\omega(\mathbf{x}, 0)$ and $c_i(\mathbf{x}, 0)$:
\begin{eqnarray}
    \omega(\mathbf{x}, 0) = 0\,, \;\; c_i(\mathbf{x}, 0) = \bar{c}_i + a_i\, \eta_i(\mathbf{x}),\label{eq:init}
\end{eqnarray}
where $\bar{c}_i$, $\eta_i(\mathbf{x})$, and $a_i$ are, respectively, the average concentration, spatial white noise uniformly distributed in the interval $[0, 1]$, and noise amplitude for the $i_{th}$ component. We use the following fixed parameter values throughout our simulations: $\epsilon = 3 \pi/256$, $M = 10^{-4}, \Lambda = 6$, $a_i = 10^{-3}$. To resolve the interfaces, we ensure at least three grid points span each interfacial region. 

\section{Results} \label{sec…3}
\begin{figure*}
{
\includegraphics[width=\linewidth]{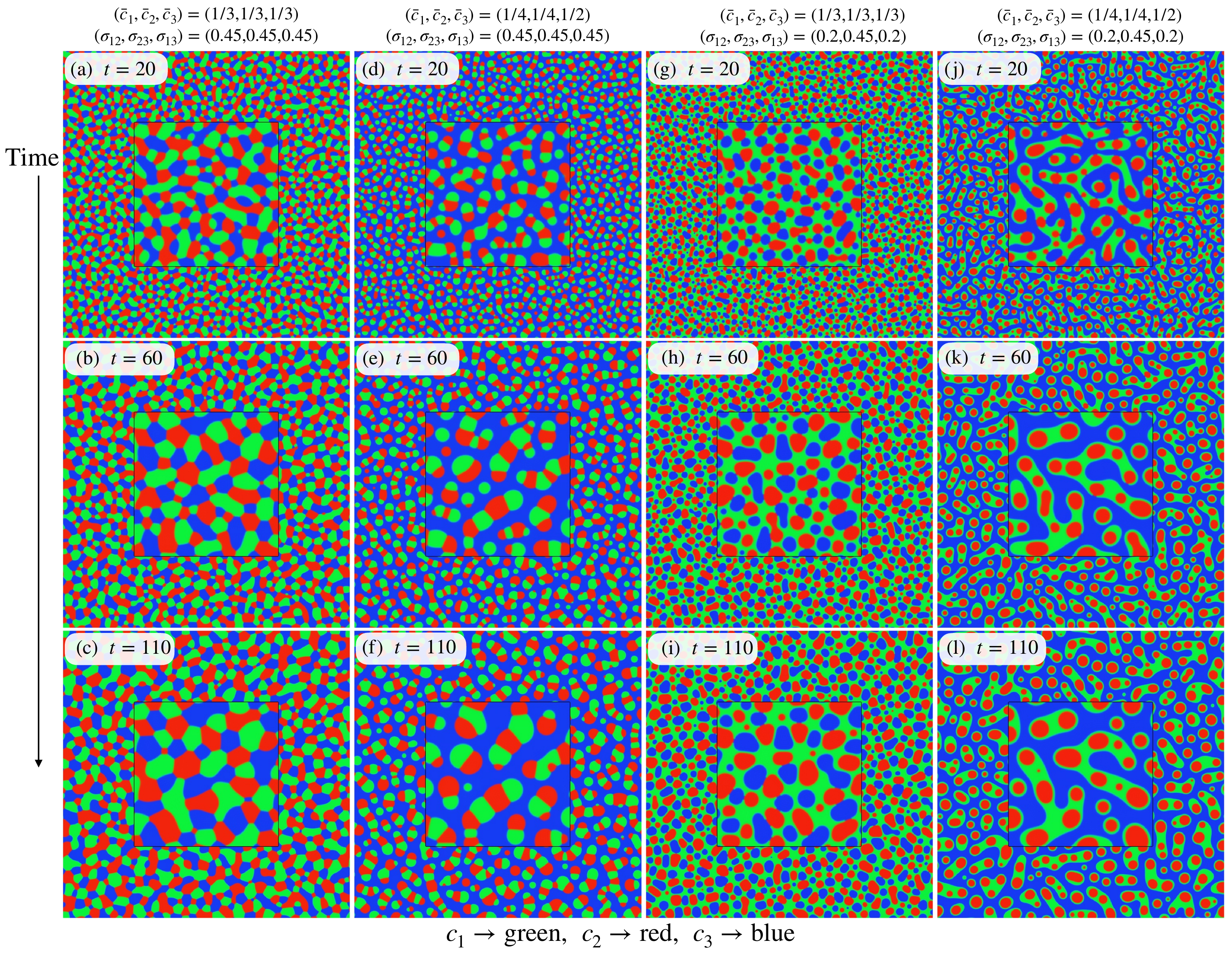}
}
\caption{\label{fig:color1} 
{
(Case I: Ternary CH model) 
Pseudocolor plots of the concentration fields $(c_1, c_2, c_3)$, shown in green, red, and blue respectively to distinguish the three components, at times $t = 20, 60, 110$ (indicated by the downward arrow showing time evolution). The plots are overlaid with magnified views of their central regions. Panels (a)--(c): $(\tilde{c}_1, \tilde{c}_2, \tilde{c}_3) = (1/3, 1/3, 1/3)$, $(\sigma_{12}, \sigma_{23}, \sigma_{13}) = (0.45, 0.45, 0.45)$; (d)--(f): $(\tilde{c}_1, \tilde{c}_2, \tilde{c}_3) = (1/4, 1/4, 1/2)$, $(\sigma_{12}, \sigma_{23}, \sigma_{13}) = (0.45, 0.45, 0.45)$;
(g)--(i): $(\tilde{c}_1, \tilde{c}_2, \tilde{c}_3) = (1/3, 1/3, 1/3)$, $(\sigma_{12}, \sigma_{23}, \sigma_{13}) = (0.2, 0.45, 0.2)$; (j)--(l): $(\tilde{c}_1, \tilde{c}_2, \tilde{c}_3) = (1/4, 1/4, 1/2)$, $(\sigma_{12}, \sigma_{23}, \sigma_{13}) = (0.2, 0.45, 0.2)$. Panels (a)--(f) show partial wetting with $W_i < 0$ for all $i \in \{1,2,3\}$, whereas (g)--(l) show complete wetting, where the green phase wets the red and blue phases with $W_1 > 0$, $W_2, W_3 < 0$.}
}
\end{figure*}
\begin{figure*}
{
\includegraphics[width=\linewidth]{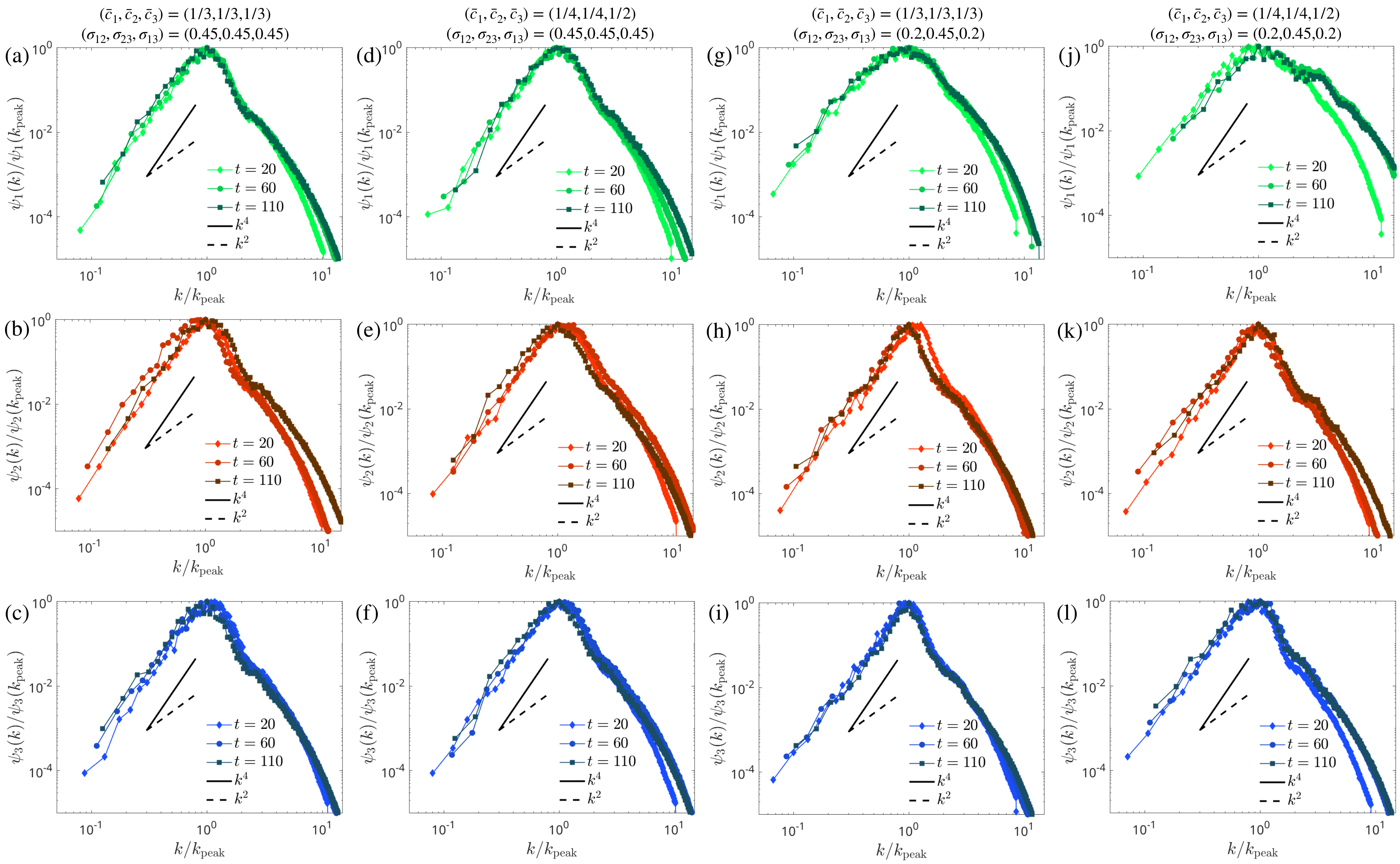}
}
\caption{\label{fig:spectra1}(Case I: Ternary CH model) Log-log plots of the spectral densities $\psi_1(k)$, $\psi_2(k)$, and $\psi_3(k)$, shown in green, red, and blue, respectively. The color coding corresponds to the pseudocolor plots in Fig.~\ref{fig:color1}. Both axes are normalized by $\psi_i(k_{\text{peak}})$ and $k_{\text{peak}}$, respectively, where $k_{\text{peak}}$ is the wavenumber at which the spectral density attains its maximum. Panels (a)--(c): $(\tilde{c}_1, \tilde{c}_2, \tilde{c}_3) = (1/3, 1/3, 1/3)$, $(\sigma_{12}, \sigma_{23}, \sigma_{13}) = (0.45, 0.45, 0.45)$; (d)--(f): $(\tilde{c}_1, \tilde{c}_2, \tilde{c}_3) = (1/4, 1/4, 1/2)$, $(\sigma_{12}, \sigma_{23}, \sigma_{13}) = (0.45, 0.45, 0.45)$; (g)--(i): $(\tilde{c}_1, \tilde{c}_2, \tilde{c}_3) = (1/3, 1/3, 1/3)$, $(\sigma_{12}, \sigma_{23}, \sigma_{13}) = (0.2, 0.45, 0.2)$; (j)--(l): $(\tilde{c}_1, \tilde{c}_2, \tilde{c}_3) = (1/4, 1/4, 1/2)$, $(\sigma_{12}, \sigma_{23}, \sigma_{13}) = (0.2, 0.45, 0.2)$. The spectral densities are evaluated from the concentration fields $(c_1, c_2, c_3)$ shown in Fig.~\ref{fig:color1}. In the low wavenumber limit they exhibit power-law behavior indicated by $k^2$ (black solid line) and $k^4$ (black dotted line).
}
\end{figure*}
The initially homogeneous mixture, specified by the concentration fields in Eq.~\eqref{eq:init}, becomes unstable and undergoes phase separation into distinct pure phases separated by diffuse interfaces. While the underlying dynamics minimize the free energy and would eventually lead to a fully phase-separated equilibrium state, our interest lies in the intermediate stages of this process. As we show below, these transient states display complex morphologies and striking signatures of hyperuniformity. Hyperuniformity for each component $c_i$ is characterized by the spectral densities $\psi_i(k,t)$ defined as 
\begin{eqnarray}
    \psi_i(k, t) &=& \frac{S^c_i(k, t)}{2\pi k}\\
     S^c_i(k, t) &=& \displaystyle \sum_{k\leq|\mathbf k'|<k+1} |\hat c_i(\mathbf k', t)|^2 
\end{eqnarray}
for $i=1,2,3$. If $\psi_i(k,t) \to 0$ according to some power law behavior $\psi_i(k,t) \sim k^\alpha$, with $\alpha > 0$, as the wave number $k = |\mathbf{k}| \to 0$, systems are called hyperuniform with different classifications depending on $\alpha$, e.g. class-I hyperuniformity for $\alpha > 1$ \cite{Tor18} . For the binary CH equations the scaling of the characteristic length scale $L(t) \sim t ^{1/3}$ translates into the scaling of the spectral density and leads to $\alpha = 4$ \cite{Tomita1991,de2024hyperuniformity}. For ternary systems and systems under the influence of hydrodynaics such analytical results do not exist. For the binary case it has been numerically demonstrated that the scaling with $\alpha \approx 4$ remains even if hydrodynamic interactions in a viscous regime are present \cite{padhan2025suppression}. Additional computations (not shown) demonstrate that the exponent decreases in the inertial regime. 

In order to explore the impact of hydrodynamics in ternary systems we organize our results into three cases: \\
(i) {\bf Case I}: ternary phase separation without hydrodynamics, governed by the ternary CH equations; \\ 
(ii) {\bf Case II}: ternary phase separation with viscous hydrodynamics, governed by the ternary CHNS equations at high viscosity; \\
(iii) {\bf Case III}: ternary phase separation with inertial hydrodynamics, governed by the ternary CHNS equations at low viscosity. \\
Each of these cases is further explored under four parameter regimes that yield distinct morphologies: \\(1) \textbf{Sym--Sym}: symmetric concentrations and symmetric surface tensions, with $(c_1,c_2,c_3) = (1/3,1/3,1/3)$ and $(\sigma_{12}, \sigma_{23}, \sigma_{13}) = (0.45,0.45,0.45)$; \\(2) \textbf{Asym--Sym}: asymmetric concentrations and symmetric surface tensions, with $(c_1,c_2,c_3) = (1/4,1/4,1/2)$ and $(\sigma_{12}, \sigma_{23}, \sigma_{13}) = (0.45,0.45,0.45)$;\\ (3) \textbf{Sym--Asym}: symmetric concentrations and asymmetric surface tensions, with $(c_1,c_2,c_3) = (1/3,1/3,1/3)$ and $(\sigma_{12}, \sigma_{23}, \sigma_{13}) = (0.2,0.45,0.2)$;\\(4) \textbf{Asym--Asym}: asymmetric concentrations and asymmetric surface tensions, with $(c_1,c_2,c_3) = (1/4,1/4,1/2)$ and $(\sigma_{12}, \sigma_{23}, \sigma_{13}) = (0.2,0.45,0.2)$. \\
Each setting is considered over time and shown at three representative time instances, which are chosen to roughly double the characteristic length scale $L$ and being consistent between the different cases. We analyze $\psi_i(k,t)$ as $k \to 0$ for each of these time instances.
\subsection{Case I}
\begin{figure*}
{
\includegraphics[width=\linewidth]{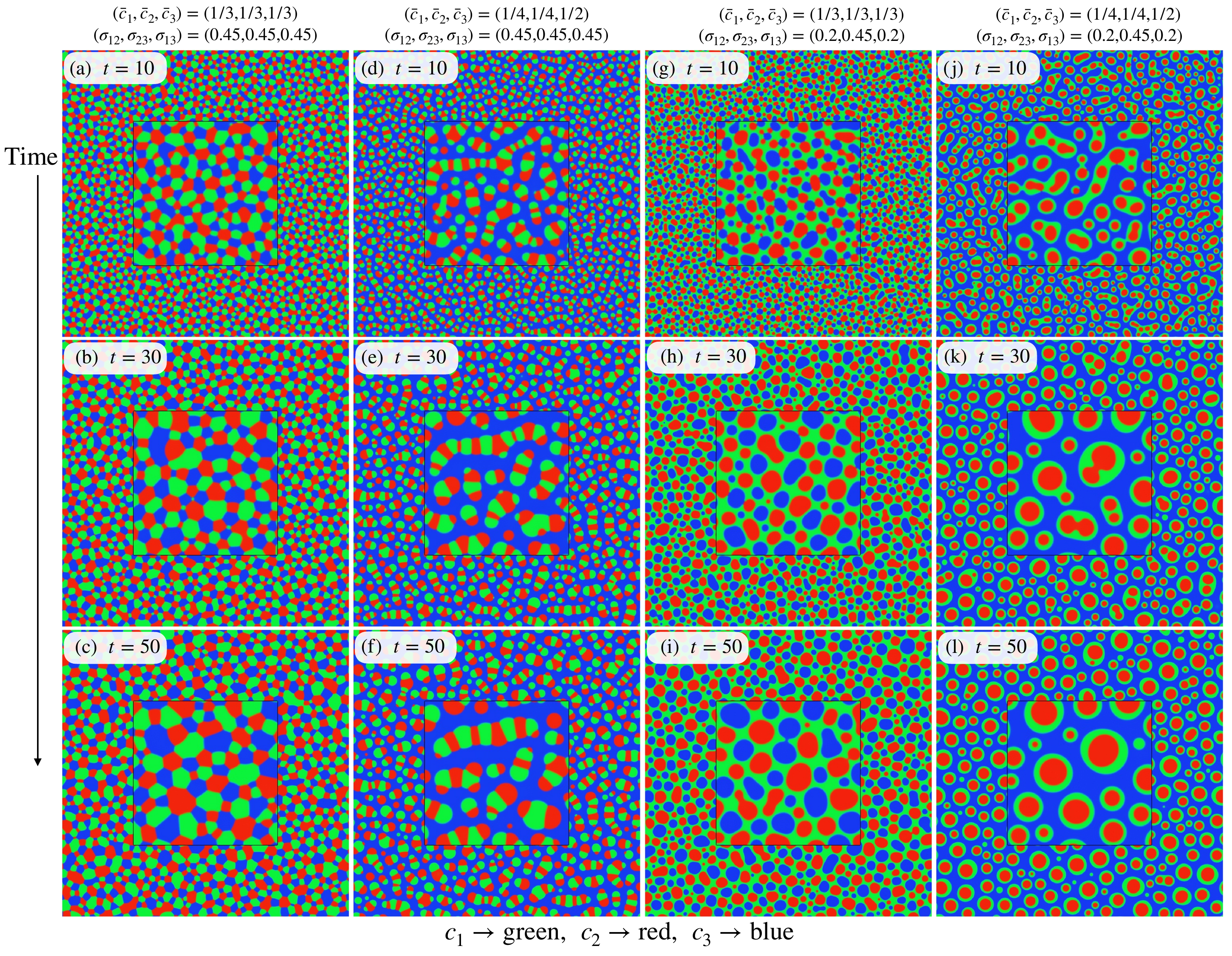}
\caption{\label{fig:color2}
(Case II: Ternary CHNS model with high viscosity $\nu = 1.5$) 
Pseudocolor plots of the concentration fields $(c_1, c_2, c_3)$, shown in green, red, and blue respectively to distinguish the three components, at times $t = 10, 30, 50$ (indicated by the downward arrow showing time evolution). The plots are overlaid with magnified views of their central regions. Panels (a)--(c): $(\tilde{c}_1, \tilde{c}_2, \tilde{c}_3) = (1/3, 1/3, 1/3)$, $(\sigma_{12}, \sigma_{23}, \sigma_{13}) = (0.45, 0.45, 0.45)$; (d)--(f): $(\tilde{c}_1, \tilde{c}_2, \tilde{c}_3) = (1/4, 1/4, 1/2)$, $(\sigma_{12}, \sigma_{23}, \sigma_{13}) = (0.45, 0.45, 0.45)$;
(g)--(i): $(\tilde{c}_1, \tilde{c}_2, \tilde{c}_3) = (1/3, 1/3, 1/3)$, $(\sigma_{12}, \sigma_{23}, \sigma_{13}) = (0.2, 0.45, 0.2)$; (j)--(l): $(\tilde{c}_1, \tilde{c}_2, \tilde{c}_3) = (1/4, 1/4, 1/2)$, $(\sigma_{12}, \sigma_{23}, \sigma_{13}) = (0.2, 0.45, 0.2)$. Panels (a)--(f) show partial wetting with $W_i < 0$ for all $i \in \{1,2,3\}$, whereas (g)--(l) show complete wetting, where the green phase wets the red and blue phases with $W_1 > 0$, $W_2, W_3 < 0$.
}
}
\end{figure*}
In Fig.~\ref{fig:color1}, we show the temporal evolution of the pseudocolor plots of the concentration fields $(c_1, c_2, c_3)$, where $c_1$, $c_2$, and $c_3$ are represented by green, red, and blue, respectively. The plots reveal the progressive domain growth of the three distinct phases as the system evolves. The four columns correspond to the four parameter regimes introduced earlier: column 1 [(a)–(c)] shows the \textbf{Sym--Sym} case, column 2 [(d)–(f)] the \textbf{Asym--Sym} case, column 3 [(g)–(i)] the \textbf{Sym--Asym} case, and column 4 [(j)–(l)] the \textbf{Asym--Asym} case. The different morphologies become appearent. In the \textbf{Sym--Sym} case, the symmetry among the three components leads to symmetric tilings with interfaces connected by triple junctions. The symmetry between the three phases is lost as soon as the composition or the surface tension becomes asymemtric. In the \textbf{Asym--Sym} case, the system forms connected droplet-like domains arranged in repeating patterns, where droplets of one phase alternate with those of another, so-called partial engulfment. These morphologies are consistent with partial wetting, as all three surface tension coefficients are equal, $\sigma_{12} = \sigma_{23} = \sigma_{13}$, giving wetting parameters $W_1 = W_2 = W_3$. In this regime, we also observe the formation of Janus droplets consisting of $c_1$ (green) and $c_2$ (red) components and also single green and red droplets. Additionally, we find compound droplets in which a larger droplet of one phase (e.g., green) bridges two smaller droplets of another (e.g., red), resembling triplet structures. The remaining two cases, \textbf{Sym--Asym} and \textbf{Asym--Asym}, correspond to the complete wetting regime, where $\sigma_{23} > \sigma_{12} + \sigma_{13}$, leading to $W_1 > 0$, while $W_2, W_3 < 0$. In these cases, the green phase preferentially wets the red and blue phases. In the \textbf{Sym--Asym} case, this results in red and blue droplets being separated by a green intermediate layer, as the system seeks to minimize the high energetic cost of forming direct $23$-interfaces. This behavior is consistent with the free energy landscape shown in Fig.~\ref{fig:fr_en}(b). The \textbf{Asym--Asym} case leads to the formation of double emulsion structures, in which red droplets are engulfed within green droplets, all embedded in the blue phase. In some instances, more complex emulsions emerge, with multiple red droplets embedded within a single green domain. Such structures coarsen over time, typically leading to one circular red droplet within a larger green droplet. Within the overall coarsening process first the green droplets merge and within the newly formed larger green droplet the red droplets merge.

The corresponding plots for the spectral densities $\psi_i(k,t)$ are shown in Fig.~\ref{fig:spectra1}. Using the same color coding they are shown separately for the green ($c_1$), red ($c_2$) and blue ($c_3$) components, for the same time instances as considered in Fig.~\ref{fig:color1}. In the low-wavenumber limit, the spectral densities show power laws, whose exponents $\alpha_i(t)$ are extracted using linear regression fit. The time evolution of these exponents is shown in Fig.~\ref{fig:exponents} (a)-(d).


\subsection{Case II}
\begin{figure*}
{
\includegraphics[width=\linewidth]{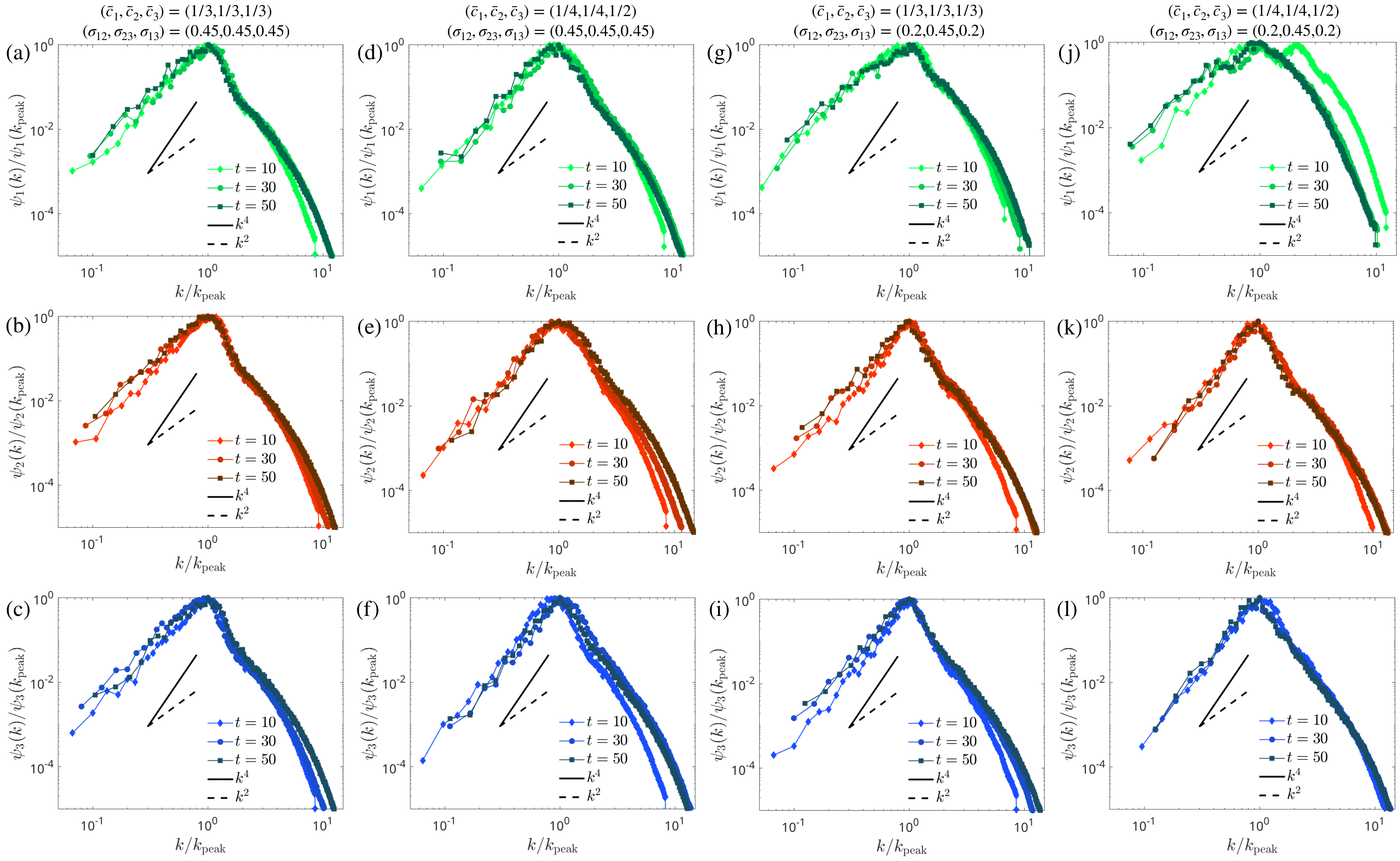}
} 
\caption{\label{fig:spectra2}(Case II: Ternary CHNS model with high viscosity $\nu = 1.5$) Log-log plots of the spectral densities $\psi_1(k)$, $\psi_2(k)$, and $\psi_3(k)$, shown in green, red, and blue, respectively. The color coding corresponds to the pseudocolor plots in Fig.~\ref{fig:color2}. Both axes are normalized by $\psi_i(k_{\text{peak}})$ and $k_{\text{peak}}$, respectively, where $k_{\text{peak}}$ is the wavenumber at which the spectral density attains its maximum. Panels (a)--(c): $(\tilde{c}_1, \tilde{c}_2, \tilde{c}_3) = (1/3, 1/3, 1/3)$, $(\sigma_{12}, \sigma_{23}, \sigma_{13}) = (0.45, 0.45, 0.45)$; (d)--(f): $(\tilde{c}_1, \tilde{c}_2, \tilde{c}_3) = (1/4, 1/4, 1/2)$, $(\sigma_{12}, \sigma_{23}, \sigma_{13}) = (0.45, 0.45, 0.45)$; (g)--(i): $(\tilde{c}_1, \tilde{c}_2, \tilde{c}_3) = (1/3, 1/3, 1/3)$, $(\sigma_{12}, \sigma_{23}, \sigma_{13}) = (0.2, 0.45, 0.2)$; (j)--(l): $(\tilde{c}_1, \tilde{c}_2, \tilde{c}_3) = (1/4, 1/4, 1/2)$, $(\sigma_{12}, \sigma_{23}, \sigma_{13}) = (0.2, 0.45, 0.2)$. The spectral densities are evaluated from the concentration fields $(c_1, c_2, c_3)$ shown in Fig.~\ref{fig:color2}. In the low wavenumber limit they exhibit power-law behavior indicated by $k^2$ (black solid line) and $k^4$ (black dotted line).
}
\end{figure*}
We consider the same investigations but now under hydrodynamic interactions and high viscosity. In Fig. \ref{fig:color2} we show the morphologies at different time instances and in Fig. \ref{fig:spectra2} the corresponding spectral densities $\psi_i(k,t)$.  The extracted exponents in the low wavenumber limit are shown in Fig.~\ref{fig:exponents} (e)-(h). The selected time instances differ due to the increased coarsening resulting from the hydrodynamic interactions in the viscous regime. This also has a consequence on the observed morphologies. While the \textbf{Sym-Sym}, \textbf{Asym-Sym} and \textbf{Sym-Asym} cases the morphologies are similar to Case I, the chain-like structures of $c_1$ and $c_2$ in the \textbf{Asym--Asym} case appearent for all considered times in Fig. \ref{fig:color1} (j)-(l), are only present at early stages. At late stages the morphology is dominated by full engulfment of the red component and a single red droplet within each green droplet. 
\begin{figure*}
{
\includegraphics[width=\linewidth]{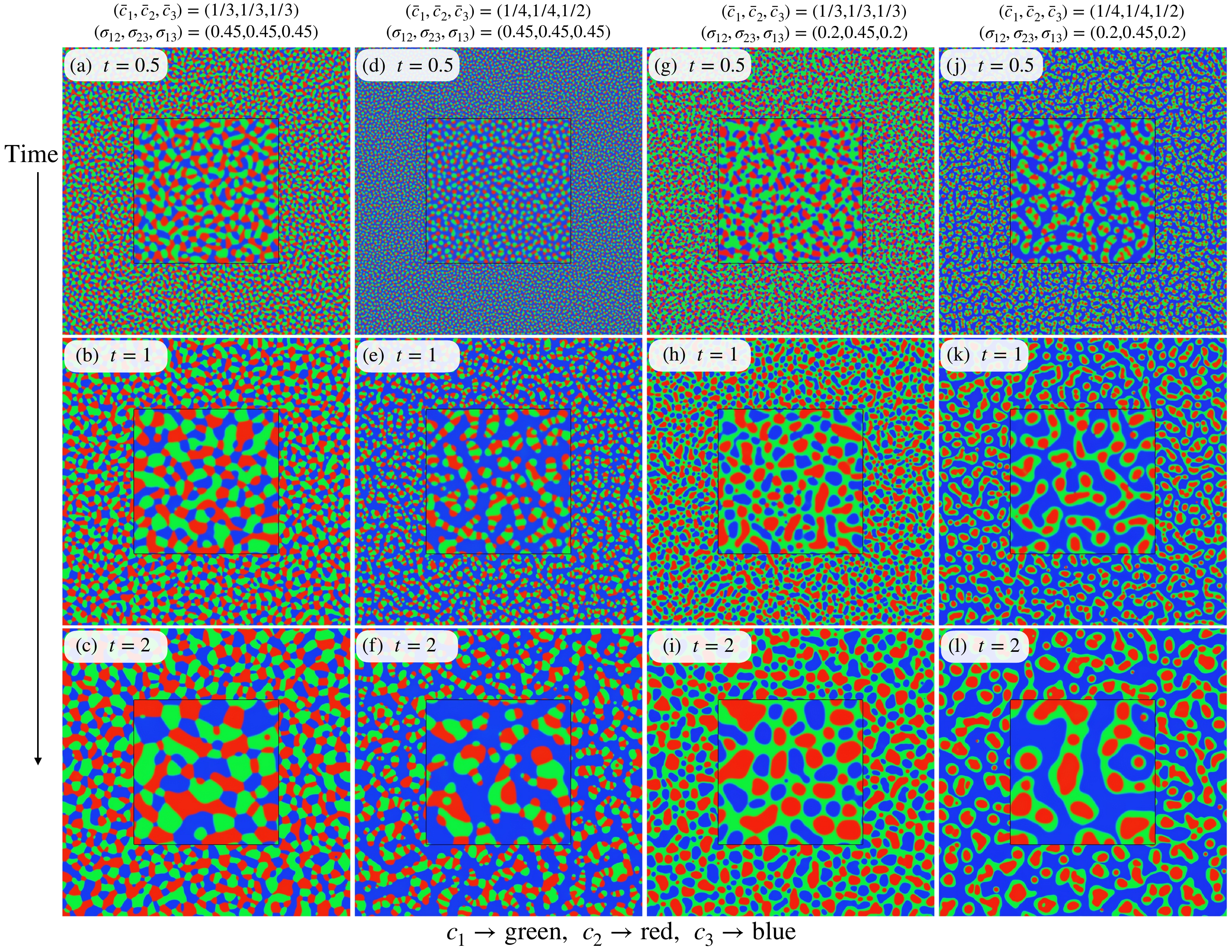}
\caption{\label{fig:color3}
(Case III: Ternary CHNS model with low viscosity $\nu = 0.01$) 
Pseudocolor plots of the concentration fields $(c_1, c_2, c_3)$, shown in green, red, and blue respectively to distinguish the three components, at times $t = 0.5, 1, 2$ (indicated by the downward arrow showing time evolution). The plots are overlaid with magnified views of their central regions. Panels (a)--(c): $(\tilde{c}_1, \tilde{c}_2, \tilde{c}_3) = (1/3, 1/3, 1/3)$, $(\sigma_{12}, \sigma_{23}, \sigma_{13}) = (0.45, 0.45, 0.45)$; (d)--(f): $(\tilde{c}_1, \tilde{c}_2, \tilde{c}_3) = (1/4, 1/4, 1/2)$, $(\sigma_{12}, \sigma_{23}, \sigma_{13}) = (0.45, 0.45, 0.45)$;
(g)--(i): $(\tilde{c}_1, \tilde{c}_2, \tilde{c}_3) = (1/3, 1/3, 1/3)$, $(\sigma_{12}, \sigma_{23}, \sigma_{13}) = (0.2, 0.45, 0.2)$; (j)--(l): $(\tilde{c}_1, \tilde{c}_2, \tilde{c}_3) = (1/4, 1/4, 1/2)$, $(\sigma_{12}, \sigma_{23}, \sigma_{13}) = (0.2, 0.45, 0.2)$. Panels (a)--(f) show partial wetting with $W_i < 0$ for all $i \in \{1,2,3\}$, whereas (g)--(l) show complete wetting, where the green phase wets the red and blue phases with $W_1 > 0$, $W_2, W_3 < 0$. Inertia drives large-scale fluctuations in the phase separation dynamics, which are absent in cases I and II.}
}
\end{figure*}
\begin{figure*}
{
\includegraphics[width=\linewidth]{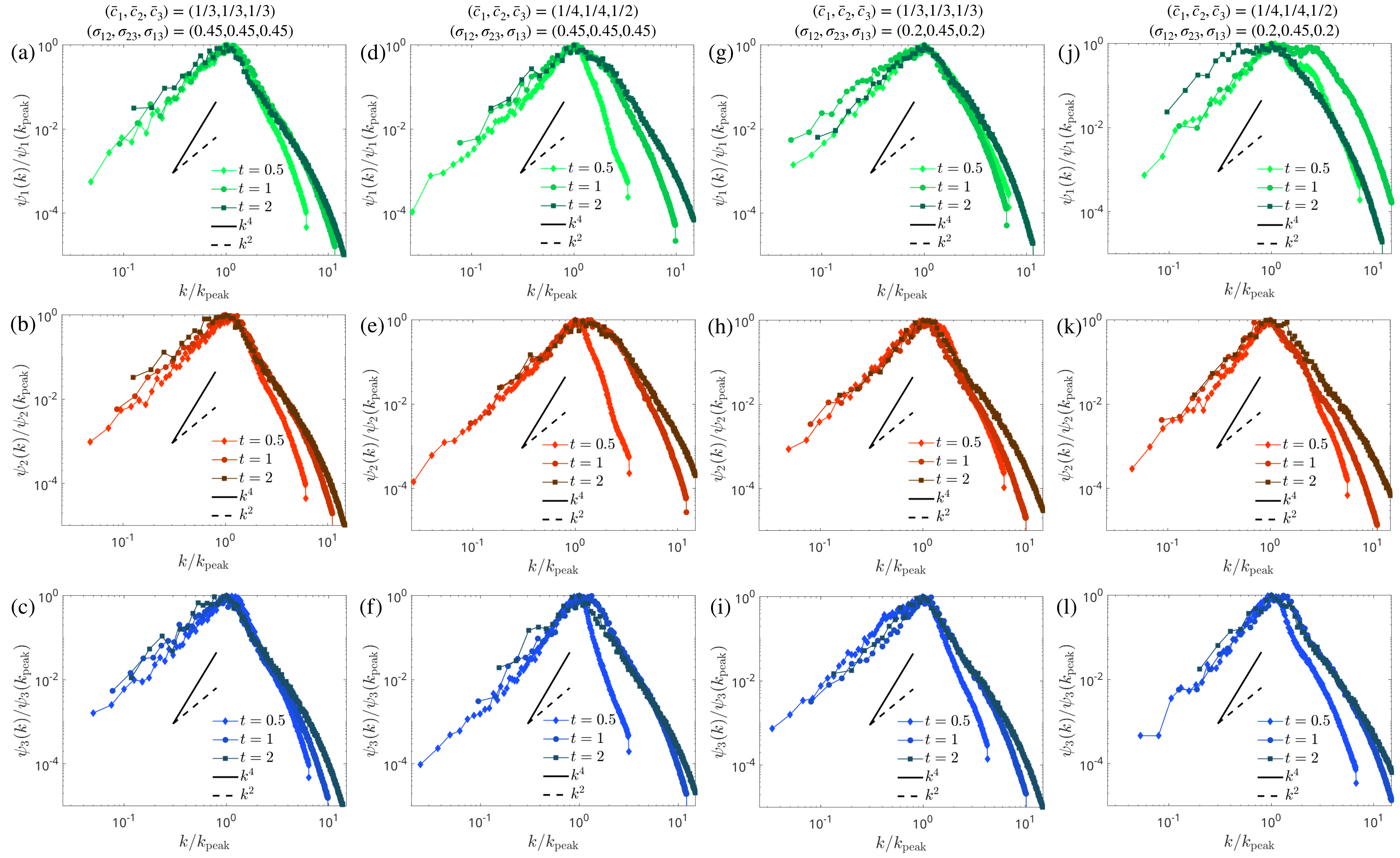}
}
\caption{\label{fig:spectra3}(Case III: Ternary CHNS model with low viscosity $\nu = 0.01$) Log-log plots of the spectral densities $\psi_1(k)$, $\psi_2(k)$, and $\psi_3(k)$, shown in green, red, and blue, respectively. The color coding corresponds to the pseudocolor plots in Fig.~\ref{fig:color3}. Both axes are normalized by $\psi_i(k_{\text{peak}})$ and $k_{\text{peak}}$, respectively, where $k_{\text{peak}}$ is the wavenumber at which the spectral density attains its maximum. Panels (a)--(c): $(\tilde{c}_1, \tilde{c}_2, \tilde{c}_3) = (1/3, 1/3, 1/3)$, $(\sigma_{12}, \sigma_{23}, \sigma_{13}) = (0.45, 0.45, 0.45)$; (d)--(f): $(\tilde{c}_1, \tilde{c}_2, \tilde{c}_3) = (1/4, 1/4, 1/2)$, $(\sigma_{12}, \sigma_{23}, \sigma_{13}) = (0.45, 0.45, 0.45)$; (g)--(i): $(\tilde{c}_1, \tilde{c}_2, \tilde{c}_3) = (1/3, 1/3, 1/3)$, $(\sigma_{12}, \sigma_{23}, \sigma_{13}) = (0.2, 0.45, 0.2)$; (j)--(l): $(\tilde{c}_1, \tilde{c}_2, \tilde{c}_3) = (1/4, 1/4, 1/2)$, $(\sigma_{12}, \sigma_{23}, \sigma_{13}) = (0.2, 0.45, 0.2)$. The spectral densities are evaluated from the concentration fields $(c_1, c_2, c_3)$ shown in Fig.~\ref{fig:color3}. In the low wavenumber limit they exhibit power-law behavior indicated by $k^2$ (black solid line) and $k^4$ (black dotted line).
}
\end{figure*}
\subsection{Case III}

We consider the same investigations but now under hydrodynamic interactions and low viscosity. In Fig. \ref{fig:color3} we show the morphologies at different time instances and in Fig. \ref{fig:spectra3} the corresponding spectral densities $\psi_i(k,t)$. The extracted exponents in the low wavenumber limit are shown in Fig.~\ref{fig:exponents}(i)-(l). The selected time instances again differ due to the further increased coarsening resulting from the hydrodynamic interactions in the inertial regime. The emerging morphologies are less regular. This is true for all four cases, \textbf{Sym-Sym}, \textbf{Asym-Sym}, \textbf{Sym-Asym} and \textbf{Asym--Asym} and a result of the inertia, which drives large-scale fluctuations. However, the basic features remain. The \textbf{Sym-Sym} case is symmetric with interfaces connected by triple junctions, the \textbf{Asym-Sym} case leads to connected droplet-like domains, reminescent of partial engulfment, the \textbf{Sym-Asym} case leads to red and blue droplets separated by a green intermediate layer, and the \textbf{Asym--Asym} case shows full engulfment or double emulsion structures. 

\subsection{Comparison}

\begin{figure*}
{
\includegraphics[width=\linewidth]{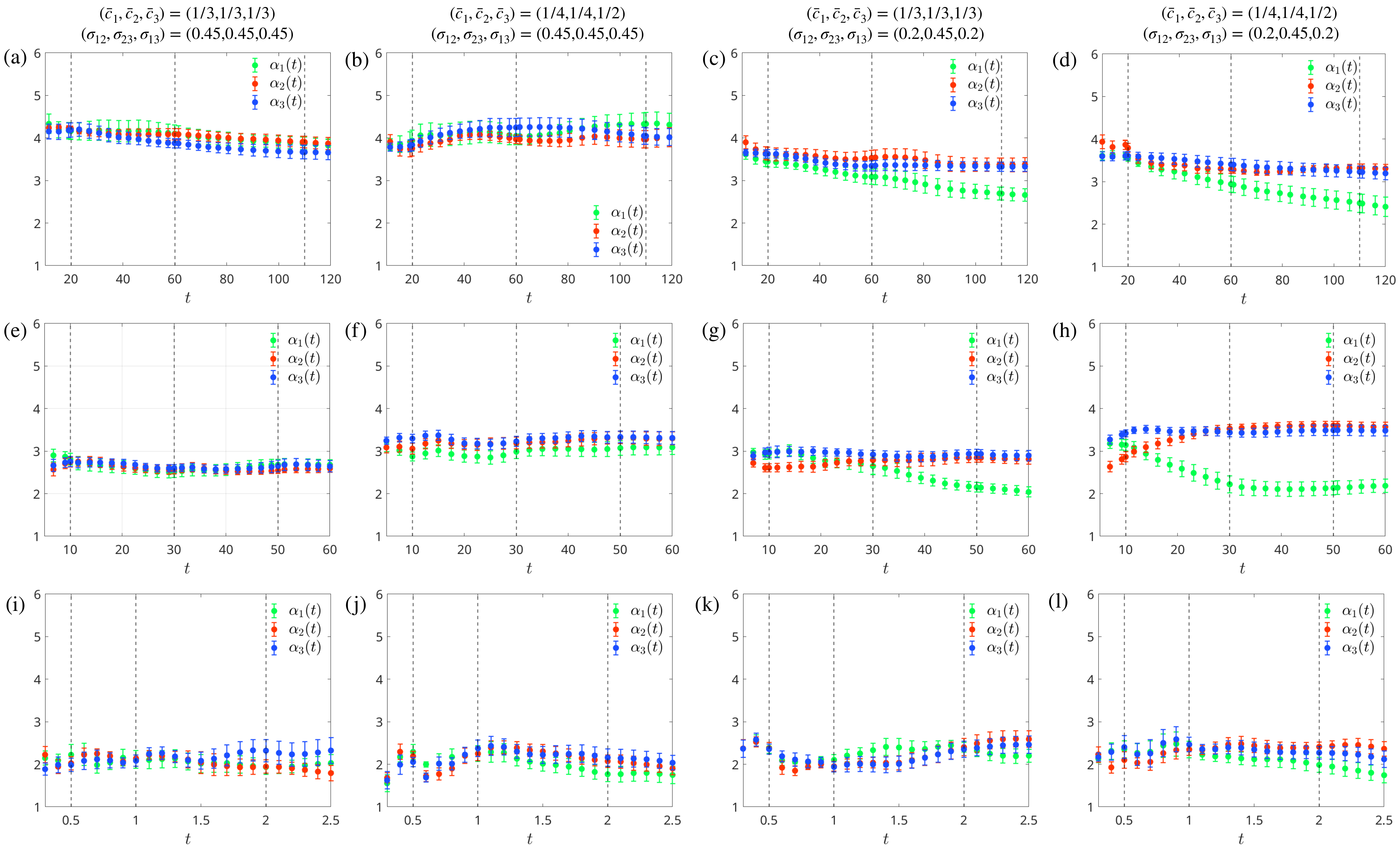}
}
\caption{\label{fig:exponents} Temporal evolution of the low-wavenumber scaling exponents $\alpha_1(t)$, $\alpha_2(t)$, and $\alpha_3(t)$ extracted from $\psi_1(k, t)$, $\psi_2(k, t)$, and $\psi_3(k, t)$, respectively, in the range $k_{\mathrm{min}} < k < k_{\mathrm{peak}}(t)$. Here, $k_{\mathrm{min}} = 2\pi / L$, with $L$ being the side length of the square simulation box. The color coding corresponds to the colors used in Figs.~(\ref{fig:color1})-(\ref{fig:spectra3}). Row-1: (Case I) ternary CH; Row-2: (Case II) ternary CHNS model with high viscosity $\nu = 1.5$); Row-3: (Case III) ternary CHNS model with low viscosity $\nu = 0.01$). Column-1: $(\tilde{c}_1, \tilde{c}_2, \tilde{c}_3) = (1/3, 1/3, 1/3)$, $(\sigma_{12}, \sigma_{23}, \sigma_{13}) = (0.45, 0.45, 0.45)$; $(\tilde{c}_1, \tilde{c}_2, \tilde{c}_3) = (1/4, 1/4, 1/2)$, $(\sigma_{12}, \sigma_{23}, \sigma_{13}) = (0.45, 0.45, 0.45)$; $(\tilde{c}_1, \tilde{c}_2, \tilde{c}_3) = (1/3, 1/3, 1/3)$, $(\sigma_{12}, \sigma_{23}, \sigma_{13}) = (0.2, 0.45, 0.2)$; (j)--(l): $(\tilde{c}_1, \tilde{c}_2, \tilde{c}_3) = (1/4, 1/4, 1/2)$, $(\sigma_{12}, \sigma_{23}, \sigma_{13}) = (0.2, 0.45, 0.2)$. The vertical dotted lines corresponds to the time instances considered in Figs. Figs.~(\ref{fig:color1})-(\ref{fig:spectra3}).}
\end{figure*}

Although essentially all considered cases display power-law scaling at the low wavenumber limit, see Figs.-\ref{fig:spectra1}, \ref{fig:spectra2} and \ref{fig:spectra3}, differences become apparent if the extracted exponents in the low wavenumber limit are compared. In Fig.~\ref{fig:exponents} the time evolution of these exponents is shown, with the selected time instances, discussed above, marked. The exponents are calculated using a linear regression fit (MATLAB's polyfit) over at least 8 data points. The error bars show the standard error of the fitted slope. Again the considered colors correspond to the color scheme used above, $c_1$ - green, $c_2$ - red and $c_3$ - blue. Several general trends can be observed: The exponents $\alpha_i(t)$ are between two and four, they only slightly vary with time (mostly decrease) within the considered time intervals, they are largest for the ternary CH model, $\alpha_i(t) \approx 4$ (Row-1) and decrease under the influence of hydrodynamics in the ternary CHNS model. The decrease depends on the viscosity, leading to $\alpha_i(t) \approx 3$ in the viscous regime (Row-2) and $\alpha_i(t) \approx 2$ in the inertial regime (Row-3). For the {\textbf{Sym-Sym}} case this is consistent, at least qualitatively, with results for the binary CH and binary CHNS models \cite{padhan2025suppression}. The behavior for the {\textbf{Asym-Sym}}
case is similar, with even slightly increased exponents, which is most prominent in Row-2. In these cases, the exponents $\alpha_1(t)$, $\alpha_2(t)$ and $\alpha_3(t)$ are also almost indistinguishable. This changes for the {\textbf{Sym-Asym}} and {\textbf{Asym-Asym}} cases. Here $c_1$ (green) differs and leads to asignificantly lower exponent $\alpha_1(t)$ as time evolves. While similar properties for $\alpha_2(t)$ (red) and $\alpha_3(t)$ (blue) for the {\textbf{Sym-Asym}} case are reasonable due to the symmetric morphologies of these components, see Figs. \ref{fig:color1}, \ref{fig:color2} and \ref{fig:color3}, these properties are at first glance  surprising for the {\textbf{Asym-Asym}} case due to the strong morphological differences between $c_2$ (red) and $c_3$ (blue). The first with single droplets fully engulfed by $c_1$ (green) and the last providing the intermediate layer for the green droplets, see Figs. \ref{fig:color1}, \ref{fig:color2} and \ref{fig:color3}. However, similar morphologies (without the green phase) also emerge in binary systems for unequal composition, leading to similar exponents. This explains the behavior and highlights the difference of the green phase. For the inertial regime (Row-3) these properties are weakest or not even visible, presumably resulting from the large-scale fluctuations. 

\begin{figure*}
{
\includegraphics[width=\linewidth]{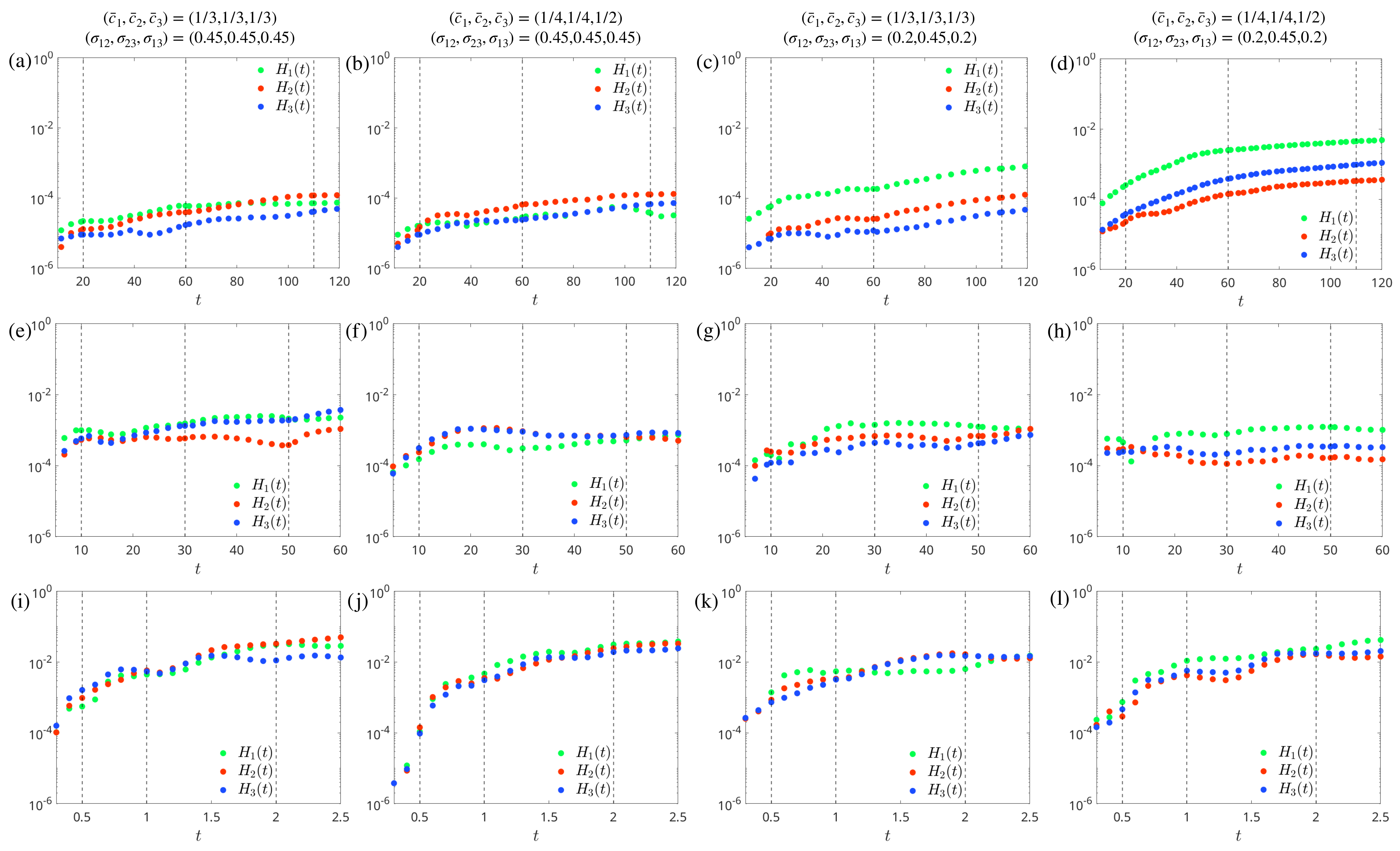}
}
\caption{\label{fig:Hyp} Temporal evolution of the hyperuniformity metrics $H_1(t)$, $H_2(t)$, and $H_3(t)$ calculated from $\psi_1(k, t)$, $\psi_2(k, t)$, and $\psi_3(k, t)$, respectively. The color coding corresponds to the colors used in Figs.~(\ref{fig:color1})-(\ref{fig:exponents}). Row-1: (Case I) ternary CH; Row-2: (Case II) ternary CHNS model with high viscosity $\nu = 1.5$); Row-3: (Case III) ternary CHNS model with low viscosity $\nu = 0.01$). Column-1: $(\tilde{c}_1, \tilde{c}_2, \tilde{c}_3) = (1/3, 1/3, 1/3)$, $(\sigma_{12}, \sigma_{23}, \sigma_{13}) = (0.45, 0.45, 0.45)$; $(\tilde{c}_1, \tilde{c}_2, \tilde{c}_3) = (1/4, 1/4, 1/2)$, $(\sigma_{12}, \sigma_{23}, \sigma_{13}) = (0.45, 0.45, 0.45)$; $(\tilde{c}_1, \tilde{c}_2, \tilde{c}_3) = (1/3, 1/3, 1/3)$, $(\sigma_{12}, \sigma_{23}, \sigma_{13}) = (0.2, 0.45, 0.2)$; (j)--(l): $(\tilde{c}_1, \tilde{c}_2, \tilde{c}_3) = (1/4, 1/4, 1/2)$, $(\sigma_{12}, \sigma_{23}, \sigma_{13}) = (0.2, 0.45, 0.2)$. The vertical dotted lines corresponds to the time instances considered in Figs. Figs.~(\ref{fig:color1})-(\ref{fig:spectra3}).}
\end{figure*}
While according to these results all components in all considered cases lead to $\alpha_i > 1$, and thus can be classified as class-I hyperuniformity \cite{Tor18}, this classification requires the ideal case of $k \to 0$ and thus infinite systems. In order to deal with the more realistic situation of finite systems various measures have been proposed. One is the hyperuniformity metric 
\begin{equation}
    H_i = \frac{\psi_i(k_{min},t)}{\psi_i(k_{peak},t)}
\end{equation} 
where $k_{peak}$ corresponds to the peak of the spectra shown in Figs. \ref{fig:spectra1}, \ref{fig:spectra2} and \ref{fig:spectra3} and $k_{min}$ to the lowest wave number considered in the simulations. For an ideal hyperuniform system we have
$H_i = 0$. However, $H_i < 10^{-4}$ is typically considered as effectively
hyperuniform and $H_i < 10^{-2}$ as nearly hyperuniform \cite{Tor18}. As our axis are normalized, $\psi_i(k_{peak}, t) = 1.0$. To compute $H_i$ at each time instance we just need to consider $\psi_i(k_{min}, t)$. We show the evolution
$H_i(t)$ for $i = 1, 2, 3$ in Fig. \ref{fig:Hyp}, again using the same color scheme as throughout the paper. Several general trends can be observed: The hyperuniformity metrics $H_i(t)$ are lowest for the ternary CH model (Row-1), with $H_i(t) < 10^{-4}$ for the \textbf{Sym-Sym} and \textbf{Asym-Sym} cases. For the \textbf{Sym-Asym} case $10^{-4} < H_1(t) < 10^{-2}$ (green), while $H_2(t) < 10^{-4}$ (red) and $H_3(t) < 10^{-4}$ (blue) and for the \textbf{Asym-Asym} case  $10^{-4} < H_1(t) < 10^{-2}$ (green) and $10^{-4} < H_3(t) < 10^{-2}$ (blue) and only $H_3(t) < 10^{-4}$ (red). Increased asymmetry and especially complete wetting thus weakens hyperuniformity from effectively hyperuniform to nearly hyperuniform. In the \textbf{Asym-Asym} case only component $c_3$ (red), which is characterized by engulfed droplets remains effectively hyperuniform. Consistently, the phase which completely wets the other two phases, here $c_1$ (green) shows the weakest properties of hyperuniformity. This corresponds also to the lowest exponent $\alpha_i$ for this phase. Under the influence of hydrodynamics in the viscous regime (Row-2) this asymmetry between the different components is still present but less pronounced. For the ternary CHNS equations with high viscosity the hyperuniformity metrics $H_i(t)$ increases for all components. We have $10^{-4} < H_i(t) < 10^{-2}$, and thus only nearly hyperuniformity, in all cases and for all components. Only for the \textbf{Asym-Asym} case $c_3$ (red), again characterized by engulfed droplets leads to $H_3(t) \approx 10^{-4}$ and shows the strongest characteristics of hyperuniformity. In the inertia regime, the ternary CHNS equations with low viscosity (Row-3), the values further increase and even reach at late time instances a regime in which $10^{-2} < H_i(t)$, which according to this measure no longer shows characteristics of hyperuniformity.  

\section{Conclusion} \label{sec:4}
We have carried out direct numerical simulations of ternary phase-field models consisting of three concentration fields. We considered the ternary CH and ternary CHNS equations varying the kinetic viscocity $\nu$, the initial compositions $(c_1, c_2, c_3)$ and the interfacial tensions $(\sigma_{12}, \sigma_{13}, \sigma_{23})$. We addressed three cases: \textbf{Case I} (ternary CH equations), \textbf{Case II} (ternary CHNS equations - viscous regime) and \textbf{Case III} (ternary CHNS equations - inertial regime), and within these cases four scenarios for the compsition and the interfacial tension which read: \textbf{Sym-Sym}, \textbf{Asym-Sym}, \textbf{Sym-Asym} and \textbf{Asym--Asym}. This leads to different morphologies which are qualitatively described and quantified with respect to characteristics of hyperuniformity for each component. 

In the asymmetric settings the morphologies show all known cases of partial and complete engulfment. For partial engulfment, in our setting the \textbf{Asym-Sym} case, droplets of different composition are wetted on each other to result in a Janus droplet or generalizations of this configuration with multiple combinations of the two phases. For complete engulfment, in our setting the \textbf{Sym-Asym} and \textbf{Asym--Asym} cases, one component can spread between the droplets of the other two components, known as sorting or one droplet can completely engulf the other, resulting in a compound droplet. These phenomena are well known for immiscible fluids with different compositions and surface tensions. However, the same phenomena are also observed in cell sorting and adhesion-based self-organization in tissue patterning. Both fields are linked to each other by the differential adhesion hypothesis of M. Steinberg \cite{doi:10.1073/pnas.48.9.1577,doi:10.1126/science.137.3532.762,doi:10.1073/pnas.48.10.1769}. Only recently the connection has also been established mathematically \cite{falco2025nonlocal}. The universal phenomenon of phase separation thus not only hold in a variety of physical
settings, they are also connected and thus allowed to be explored in the setting of multiphase fluid mixtures. 

Next to this qualitative description of the morphology, we were interested in quantifying global features, especially characteristics of hyperuniformity. This is addressed by computing the spectral densities for each component $\psi_i(t)$, the low-wavenumber scaling exponents $\alpha_i(t)$ and the hyperuniformity metrixes $H_i(t)$. Hydrodynamics decreases the characteristics of hyperuniformity. For the \textbf{Sym-Sym} case this behavior is similar to the binary setting \cite{padhan2025suppression}. Adding asymmetry we found that in partial wetting regimes, all three components display comparable degrees of hyperuniformity, which again descreses under the influence of hydrodynamics. In contrast, under complete wetting conditions — where one component preferentially wets the other two — the wetting component shows a
marked reduction in hyperuniformity (in our setting the green phase). These findings
suggest that wetting asymmetry serves as a tunable control parameter for spatial order in multiphase fluid systems.

The different settings are chosen as representative examples for asymmetries. Within the mentioned constraints any other combination can be considered with the same approach. We expect the general findings to hold also in other cases and to become even stronger for multicomponent systems beyond ternary systems.

\begin{acknowledgments}
\textit{Acknowledgments:} This work was supported by the German Research Foundation (DFG) through the project "Analysing structure-property relations in equilibrium and non-equilibrium hyperuniform
systems" (project number VO 899/32-1). We further acknowledge computing resources at JSC under grant "MORPH" and at ZIH under grant WIR.
\end{acknowledgments}
\bibliography{maain}

\begin{thebibliography}{60}%
\makeatletter
\providecommand \@ifxundefined [1]{%
 \@ifx{#1\undefined}
}%
\providecommand \@ifnum [1]{%
 \ifnum #1\expandafter \@firstoftwo
 \else \expandafter \@secondoftwo
 \fi
}%
\providecommand \@ifx [1]{%
 \ifx #1\expandafter \@firstoftwo
 \else \expandafter \@secondoftwo
 \fi
}%
\providecommand \natexlab [1]{#1}%
\providecommand \enquote  [1]{``#1''}%
\providecommand \bibnamefont  [1]{#1}%
\providecommand \bibfnamefont [1]{#1}%
\providecommand \citenamefont [1]{#1}%
\providecommand \href@noop [0]{\@secondoftwo}%
\providecommand \href [0]{\begingroup \@sanitize@url \@href}%
\providecommand \@href[1]{\@@startlink{#1}\@@href}%
\providecommand \@@href[1]{\endgroup#1\@@endlink}%
\providecommand \@sanitize@url [0]{\catcode `\\12\catcode `\$12\catcode `\&12\catcode `\#12\catcode `\^12\catcode `\_12\catcode `\%12\relax}%
\providecommand \@@startlink[1]{}%
\providecommand \@@endlink[0]{}%
\providecommand \url  [0]{\begingroup\@sanitize@url \@url }%
\providecommand \@url [1]{\endgroup\@href {#1}{\urlprefix }}%
\providecommand \urlprefix  [0]{URL }%
\providecommand \Eprint [0]{\href }%
\providecommand \doibase [0]{https://doi.org/}%
\providecommand \selectlanguage [0]{\@gobble}%
\providecommand \bibinfo  [0]{\@secondoftwo}%
\providecommand \bibfield  [0]{\@secondoftwo}%
\providecommand \translation [1]{[#1]}%
\providecommand \BibitemOpen [0]{}%
\providecommand \bibitemStop [0]{}%
\providecommand \bibitemNoStop [0]{.\EOS\space}%
\providecommand \EOS [0]{\spacefactor3000\relax}%
\providecommand \BibitemShut  [1]{\csname bibitem#1\endcsname}%
\let\auto@bib@innerbib\@empty
\bibitem [{\citenamefont {Zarzar}\ \emph {et~al.}(2015)\citenamefont {Zarzar}, \citenamefont {Sresht}, \citenamefont {Sletten}, \citenamefont {Kalow}, \citenamefont {Blankschtein},\ and\ \citenamefont {Swager}}]{zarzar2015dynamically}%
  \BibitemOpen
  \bibfield  {author} {\bibinfo {author} {\bibfnamefont {L.~D.}\ \bibnamefont {Zarzar}}, \bibinfo {author} {\bibfnamefont {V.}~\bibnamefont {Sresht}}, \bibinfo {author} {\bibfnamefont {E.~M.}\ \bibnamefont {Sletten}}, \bibinfo {author} {\bibfnamefont {J.~A.}\ \bibnamefont {Kalow}}, \bibinfo {author} {\bibfnamefont {D.}~\bibnamefont {Blankschtein}},\ and\ \bibinfo {author} {\bibfnamefont {T.~M.}\ \bibnamefont {Swager}},\ }\bibfield  {title} {\bibinfo {title} {Dynamically reconfigurable complex emulsions via tunable interfacial tensions},\ }\href@noop {} {\bibfield  {journal} {\bibinfo  {journal} {Nature}\ }\textbf {\bibinfo {volume} {518}},\ \bibinfo {pages} {520} (\bibinfo {year} {2015})}\BibitemShut {NoStop}%
\bibitem [{\citenamefont {Shono}\ \emph {et~al.}(2023)\citenamefont {Shono}, \citenamefont {Honda}, \citenamefont {Yanagisawa}, \citenamefont {Yoshikawa},\ and\ \citenamefont {Shioi}}]{shono2023spontaneous}%
  \BibitemOpen
  \bibfield  {author} {\bibinfo {author} {\bibfnamefont {M.}~\bibnamefont {Shono}}, \bibinfo {author} {\bibfnamefont {G.}~\bibnamefont {Honda}}, \bibinfo {author} {\bibfnamefont {M.}~\bibnamefont {Yanagisawa}}, \bibinfo {author} {\bibfnamefont {K.}~\bibnamefont {Yoshikawa}},\ and\ \bibinfo {author} {\bibfnamefont {A.}~\bibnamefont {Shioi}},\ }\bibfield  {title} {\bibinfo {title} {Spontaneous formation of uniform cell-sized microgels through water/water phase separation},\ }\href@noop {} {\bibfield  {journal} {\bibinfo  {journal} {Small}\ }\textbf {\bibinfo {volume} {19}},\ \bibinfo {pages} {2302193} (\bibinfo {year} {2023})}\BibitemShut {NoStop}%
\bibitem [{\citenamefont {Yanagisawa}\ \emph {et~al.}(2007)\citenamefont {Yanagisawa}, \citenamefont {Imai}, \citenamefont {Masui}, \citenamefont {Komura},\ and\ \citenamefont {Ohta}}]{yanagisawa2007growth}%
  \BibitemOpen
  \bibfield  {author} {\bibinfo {author} {\bibfnamefont {M.}~\bibnamefont {Yanagisawa}}, \bibinfo {author} {\bibfnamefont {M.}~\bibnamefont {Imai}}, \bibinfo {author} {\bibfnamefont {T.}~\bibnamefont {Masui}}, \bibinfo {author} {\bibfnamefont {S.}~\bibnamefont {Komura}},\ and\ \bibinfo {author} {\bibfnamefont {T.}~\bibnamefont {Ohta}},\ }\bibfield  {title} {\bibinfo {title} {Growth dynamics of domains in ternary fluid vesicles},\ }\href@noop {} {\bibfield  {journal} {\bibinfo  {journal} {Biophysical Journal}\ }\textbf {\bibinfo {volume} {92}},\ \bibinfo {pages} {115} (\bibinfo {year} {2007})}\BibitemShut {NoStop}%
\bibitem [{\citenamefont {Cai}\ \emph {et~al.}(2021)\citenamefont {Cai}, \citenamefont {Skabeev}, \citenamefont {Morozova},\ and\ \citenamefont {Pham}}]{cai2021fluid}%
  \BibitemOpen
  \bibfield  {author} {\bibinfo {author} {\bibfnamefont {Z.}~\bibnamefont {Cai}}, \bibinfo {author} {\bibfnamefont {A.}~\bibnamefont {Skabeev}}, \bibinfo {author} {\bibfnamefont {S.}~\bibnamefont {Morozova}},\ and\ \bibinfo {author} {\bibfnamefont {J.~T.}\ \bibnamefont {Pham}},\ }\bibfield  {title} {\bibinfo {title} {Fluid separation and network deformation in wetting of soft and swollen surfaces},\ }\href@noop {} {\bibfield  {journal} {\bibinfo  {journal} {Communications Materials}\ }\textbf {\bibinfo {volume} {2}},\ \bibinfo {pages} {21} (\bibinfo {year} {2021})}\BibitemShut {NoStop}%
\bibitem [{\citenamefont {Iqbal}\ \emph {et~al.}(2013)\citenamefont {Iqbal}, \citenamefont {Baloch}, \citenamefont {Hameed},\ and\ \citenamefont {McClements}}]{iqbal2013controlling}%
  \BibitemOpen
  \bibfield  {author} {\bibinfo {author} {\bibfnamefont {S.}~\bibnamefont {Iqbal}}, \bibinfo {author} {\bibfnamefont {M.~K.}\ \bibnamefont {Baloch}}, \bibinfo {author} {\bibfnamefont {G.}~\bibnamefont {Hameed}},\ and\ \bibinfo {author} {\bibfnamefont {D.~J.}\ \bibnamefont {McClements}},\ }\bibfield  {title} {\bibinfo {title} {Controlling w/o/w multiple emulsion microstructure by osmotic swelling and internal protein gelation},\ }\href@noop {} {\bibfield  {journal} {\bibinfo  {journal} {Food Research International}\ }\textbf {\bibinfo {volume} {54}},\ \bibinfo {pages} {1613} (\bibinfo {year} {2013})}\BibitemShut {NoStop}%
\bibitem [{\citenamefont {Clegg}\ \emph {et~al.}(2016)\citenamefont {Clegg}, \citenamefont {Tavacoli},\ and\ \citenamefont {Wilde}}]{clegg2016one}%
  \BibitemOpen
  \bibfield  {author} {\bibinfo {author} {\bibfnamefont {P.~S.}\ \bibnamefont {Clegg}}, \bibinfo {author} {\bibfnamefont {J.~W.}\ \bibnamefont {Tavacoli}},\ and\ \bibinfo {author} {\bibfnamefont {P.~J.}\ \bibnamefont {Wilde}},\ }\bibfield  {title} {\bibinfo {title} {One-step production of multiple emulsions: microfluidic, polymer-stabilized and particle-stabilized approaches},\ }\href@noop {} {\bibfield  {journal} {\bibinfo  {journal} {Soft Matter}\ }\textbf {\bibinfo {volume} {12}},\ \bibinfo {pages} {998} (\bibinfo {year} {2016})}\BibitemShut {NoStop}%
\bibitem [{\citenamefont {Cates}\ and\ \citenamefont {Nardini}(2025)}]{Cates_2025}%
  \BibitemOpen
  \bibfield  {author} {\bibinfo {author} {\bibfnamefont {M.~E.}\ \bibnamefont {Cates}}\ and\ \bibinfo {author} {\bibfnamefont {C.}~\bibnamefont {Nardini}},\ }\bibfield  {title} {\bibinfo {title} {Active phase separation: new phenomenology from non-equilibrium physics},\ }\href {https://doi.org/10.1088/1361-6633/add278} {\bibfield  {journal} {\bibinfo  {journal} {Reports on Progress in Physics}\ }\textbf {\bibinfo {volume} {88}},\ \bibinfo {pages} {056601} (\bibinfo {year} {2025})}\BibitemShut {NoStop}%
\bibitem [{\citenamefont {Bhattacharyya}\ and\ \citenamefont {Yeomans}(2023)}]{bhattacharyya2023phase}%
  \BibitemOpen
  \bibfield  {author} {\bibinfo {author} {\bibfnamefont {S.}~\bibnamefont {Bhattacharyya}}\ and\ \bibinfo {author} {\bibfnamefont {J.~M.}\ \bibnamefont {Yeomans}},\ }\bibfield  {title} {\bibinfo {title} {Phase separation driven by active flows},\ }\href@noop {} {\bibfield  {journal} {\bibinfo  {journal} {Physical Review Letters}\ }\textbf {\bibinfo {volume} {130}},\ \bibinfo {pages} {238201} (\bibinfo {year} {2023})}\BibitemShut {NoStop}%
\bibitem [{\citenamefont {Zhang}\ \emph {et~al.}(2021)\citenamefont {Zhang}, \citenamefont {Alert}, \citenamefont {Yan}, \citenamefont {Wingreen},\ and\ \citenamefont {Granick}}]{zhang2021active}%
  \BibitemOpen
  \bibfield  {author} {\bibinfo {author} {\bibfnamefont {J.}~\bibnamefont {Zhang}}, \bibinfo {author} {\bibfnamefont {R.}~\bibnamefont {Alert}}, \bibinfo {author} {\bibfnamefont {J.}~\bibnamefont {Yan}}, \bibinfo {author} {\bibfnamefont {N.~S.}\ \bibnamefont {Wingreen}},\ and\ \bibinfo {author} {\bibfnamefont {S.}~\bibnamefont {Granick}},\ }\bibfield  {title} {\bibinfo {title} {Active phase separation by turning towards regions of higher density},\ }\href@noop {} {\bibfield  {journal} {\bibinfo  {journal} {Nature Physics}\ }\textbf {\bibinfo {volume} {17}},\ \bibinfo {pages} {961} (\bibinfo {year} {2021})}\BibitemShut {NoStop}%
\bibitem [{\citenamefont {Zheng}\ \emph {et~al.}(2024)\citenamefont {Zheng}, \citenamefont {Klatt},\ and\ \citenamefont {L{\"o}wen}}]{zheng2024universal}%
  \BibitemOpen
  \bibfield  {author} {\bibinfo {author} {\bibfnamefont {Y.}~\bibnamefont {Zheng}}, \bibinfo {author} {\bibfnamefont {M.~A.}\ \bibnamefont {Klatt}},\ and\ \bibinfo {author} {\bibfnamefont {H.}~\bibnamefont {L{\"o}wen}},\ }\bibfield  {title} {\bibinfo {title} {Universal hyperuniformity in active field theories},\ }\href@noop {} {\bibfield  {journal} {\bibinfo  {journal} {Physical Review Research}\ }\textbf {\bibinfo {volume} {6}},\ \bibinfo {pages} {L032056} (\bibinfo {year} {2024})}\BibitemShut {NoStop}%
\bibitem [{\citenamefont {Cates}\ \emph {et~al.}(2010)\citenamefont {Cates}, \citenamefont {Marenduzzo}, \citenamefont {Pagonabarraga},\ and\ \citenamefont {Tailleur}}]{cates2010arrested}%
  \BibitemOpen
  \bibfield  {author} {\bibinfo {author} {\bibfnamefont {M.~E.}\ \bibnamefont {Cates}}, \bibinfo {author} {\bibfnamefont {D.}~\bibnamefont {Marenduzzo}}, \bibinfo {author} {\bibfnamefont {I.}~\bibnamefont {Pagonabarraga}},\ and\ \bibinfo {author} {\bibfnamefont {J.}~\bibnamefont {Tailleur}},\ }\bibfield  {title} {\bibinfo {title} {Arrested phase separation in reproducing bacteria creates a generic route to pattern formation},\ }\href@noop {} {\bibfield  {journal} {\bibinfo  {journal} {Proceedings of the National Academy of Sciences}\ }\textbf {\bibinfo {volume} {107}},\ \bibinfo {pages} {11715} (\bibinfo {year} {2010})}\BibitemShut {NoStop}%
\bibitem [{\citenamefont {Gouveia}\ \emph {et~al.}(2022)\citenamefont {Gouveia}, \citenamefont {Kim}, \citenamefont {Shaevitz}, \citenamefont {Petry}, \citenamefont {Stone},\ and\ \citenamefont {Brangwynne}}]{gouveia2022capillary}%
  \BibitemOpen
  \bibfield  {author} {\bibinfo {author} {\bibfnamefont {B.}~\bibnamefont {Gouveia}}, \bibinfo {author} {\bibfnamefont {Y.}~\bibnamefont {Kim}}, \bibinfo {author} {\bibfnamefont {J.~W.}\ \bibnamefont {Shaevitz}}, \bibinfo {author} {\bibfnamefont {S.}~\bibnamefont {Petry}}, \bibinfo {author} {\bibfnamefont {H.~A.}\ \bibnamefont {Stone}},\ and\ \bibinfo {author} {\bibfnamefont {C.~P.}\ \bibnamefont {Brangwynne}},\ }\bibfield  {title} {\bibinfo {title} {Capillary forces generated by biomolecular condensates},\ }\href@noop {} {\bibfield  {journal} {\bibinfo  {journal} {Nature}\ }\textbf {\bibinfo {volume} {609}},\ \bibinfo {pages} {255} (\bibinfo {year} {2022})}\BibitemShut {NoStop}%
\bibitem [{\citenamefont {Wilken}\ \emph {et~al.}(2023)\citenamefont {Wilken}, \citenamefont {Chaderjian},\ and\ \citenamefont {Saleh}}]{wilken2023spatial}%
  \BibitemOpen
  \bibfield  {author} {\bibinfo {author} {\bibfnamefont {S.}~\bibnamefont {Wilken}}, \bibinfo {author} {\bibfnamefont {A.}~\bibnamefont {Chaderjian}},\ and\ \bibinfo {author} {\bibfnamefont {O.~A.}\ \bibnamefont {Saleh}},\ }\bibfield  {title} {\bibinfo {title} {Spatial organization of phase-separated dna droplets},\ }\href@noop {} {\bibfield  {journal} {\bibinfo  {journal} {Physical Review X}\ }\textbf {\bibinfo {volume} {13}},\ \bibinfo {pages} {031014} (\bibinfo {year} {2023})}\BibitemShut {NoStop}%
\bibitem [{\citenamefont {Bansal}\ \emph {et~al.}(2022)\citenamefont {Bansal}, \citenamefont {Das},\ and\ \citenamefont {Rao}}]{bansal2022active}%
  \BibitemOpen
  \bibfield  {author} {\bibinfo {author} {\bibfnamefont {A.}~\bibnamefont {Bansal}}, \bibinfo {author} {\bibfnamefont {A.}~\bibnamefont {Das}},\ and\ \bibinfo {author} {\bibfnamefont {M.}~\bibnamefont {Rao}},\ }\bibfield  {title} {\bibinfo {title} {Active segregation dynamics in the living cell},\ }\href@noop {} {\bibfield  {journal} {\bibinfo  {journal} {Indian Journal of Physics}\ }\textbf {\bibinfo {volume} {96}},\ \bibinfo {pages} {2567} (\bibinfo {year} {2022})}\BibitemShut {NoStop}%
\bibitem [{\citenamefont {Saha}\ \emph {et~al.}(2022)\citenamefont {Saha}, \citenamefont {Das}, \citenamefont {Patra}, \citenamefont {Anilkumar}, \citenamefont {Sil}, \citenamefont {Mayor},\ and\ \citenamefont {Rao}}]{saha2022active}%
  \BibitemOpen
  \bibfield  {author} {\bibinfo {author} {\bibfnamefont {S.}~\bibnamefont {Saha}}, \bibinfo {author} {\bibfnamefont {A.}~\bibnamefont {Das}}, \bibinfo {author} {\bibfnamefont {C.}~\bibnamefont {Patra}}, \bibinfo {author} {\bibfnamefont {A.~A.}\ \bibnamefont {Anilkumar}}, \bibinfo {author} {\bibfnamefont {P.}~\bibnamefont {Sil}}, \bibinfo {author} {\bibfnamefont {S.}~\bibnamefont {Mayor}},\ and\ \bibinfo {author} {\bibfnamefont {M.}~\bibnamefont {Rao}},\ }\bibfield  {title} {\bibinfo {title} {Active emulsions in living cell membranes driven by contractile stresses and transbilayer coupling},\ }\href@noop {} {\bibfield  {journal} {\bibinfo  {journal} {Proceedings of the National Academy of Sciences}\ }\textbf {\bibinfo {volume} {119}},\ \bibinfo {pages} {e2123056119} (\bibinfo {year} {2022})}\BibitemShut {NoStop}%
\bibitem [{\citenamefont {Kaur}\ \emph {et~al.}(2021)\citenamefont {Kaur}, \citenamefont {Raju}, \citenamefont {Alshareedah}, \citenamefont {Davis}, \citenamefont {Potoyan},\ and\ \citenamefont {Banerjee}}]{kaur2021sequence}%
  \BibitemOpen
  \bibfield  {author} {\bibinfo {author} {\bibfnamefont {T.}~\bibnamefont {Kaur}}, \bibinfo {author} {\bibfnamefont {M.}~\bibnamefont {Raju}}, \bibinfo {author} {\bibfnamefont {I.}~\bibnamefont {Alshareedah}}, \bibinfo {author} {\bibfnamefont {R.~B.}\ \bibnamefont {Davis}}, \bibinfo {author} {\bibfnamefont {D.~A.}\ \bibnamefont {Potoyan}},\ and\ \bibinfo {author} {\bibfnamefont {P.~R.}\ \bibnamefont {Banerjee}},\ }\bibfield  {title} {\bibinfo {title} {Sequence-encoded and composition-dependent protein-rna interactions control multiphasic condensate morphologies},\ }\href@noop {} {\bibfield  {journal} {\bibinfo  {journal} {Nature Communications}\ }\textbf {\bibinfo {volume} {12}},\ \bibinfo {pages} {872} (\bibinfo {year} {2021})}\BibitemShut {NoStop}%
\bibitem [{\citenamefont {Hyman}\ \emph {et~al.}(2014)\citenamefont {Hyman}, \citenamefont {Weber},\ and\ \citenamefont {J{\"u}licher}}]{hyman2014liquid}%
  \BibitemOpen
  \bibfield  {author} {\bibinfo {author} {\bibfnamefont {A.~A.}\ \bibnamefont {Hyman}}, \bibinfo {author} {\bibfnamefont {C.~A.}\ \bibnamefont {Weber}},\ and\ \bibinfo {author} {\bibfnamefont {F.}~\bibnamefont {J{\"u}licher}},\ }\bibfield  {title} {\bibinfo {title} {Liquid-liquid phase separation in biology},\ }\href@noop {} {\bibfield  {journal} {\bibinfo  {journal} {Annual Review of Cell and Developmental Biology}\ }\textbf {\bibinfo {volume} {30}},\ \bibinfo {pages} {39} (\bibinfo {year} {2014})}\BibitemShut {NoStop}%
\bibitem [{\citenamefont {Balasubramaniam}\ \emph {et~al.}(2021)\citenamefont {Balasubramaniam}, \citenamefont {Doostmohammadi}, \citenamefont {Saw}, \citenamefont {Narayana}, \citenamefont {Mueller}, \citenamefont {Dang}, \citenamefont {Thomas}, \citenamefont {Gupta}, \citenamefont {Sonam}, \citenamefont {Yap} \emph {et~al.}}]{balasubramaniam2021investigating}%
  \BibitemOpen
  \bibfield  {author} {\bibinfo {author} {\bibfnamefont {L.}~\bibnamefont {Balasubramaniam}}, \bibinfo {author} {\bibfnamefont {A.}~\bibnamefont {Doostmohammadi}}, \bibinfo {author} {\bibfnamefont {T.~B.}\ \bibnamefont {Saw}}, \bibinfo {author} {\bibfnamefont {G.~H. N.~S.}\ \bibnamefont {Narayana}}, \bibinfo {author} {\bibfnamefont {R.}~\bibnamefont {Mueller}}, \bibinfo {author} {\bibfnamefont {T.}~\bibnamefont {Dang}}, \bibinfo {author} {\bibfnamefont {M.}~\bibnamefont {Thomas}}, \bibinfo {author} {\bibfnamefont {S.}~\bibnamefont {Gupta}}, \bibinfo {author} {\bibfnamefont {S.}~\bibnamefont {Sonam}}, \bibinfo {author} {\bibfnamefont {A.~S.}\ \bibnamefont {Yap}}, \emph {et~al.},\ }\bibfield  {title} {\bibinfo {title} {Investigating the nature of active forces in tissues reveals how contractile cells can form extensile monolayers},\ }\href@noop {} {\bibfield  {journal} {\bibinfo  {journal} {Nature Materials}\ }\textbf {\bibinfo {volume} {20}},\ \bibinfo {pages} {1156} (\bibinfo {year} {2021})}\BibitemShut
  {NoStop}%
\bibitem [{\citenamefont {Bray}(1994)}]{bray1994theory}%
  \BibitemOpen
  \bibfield  {author} {\bibinfo {author} {\bibfnamefont {A.~J.}\ \bibnamefont {Bray}},\ }\bibfield  {title} {\bibinfo {title} {Theory of phase-ordering kinetics},\ }\href@noop {} {\bibfield  {journal} {\bibinfo  {journal} {Advances in Physics}\ }\textbf {\bibinfo {volume} {43}},\ \bibinfo {pages} {357} (\bibinfo {year} {1994})}\BibitemShut {NoStop}%
\bibitem [{\citenamefont {Chaikin}\ \emph {et~al.}(1995)\citenamefont {Chaikin}, \citenamefont {Lubensky},\ and\ \citenamefont {Witten}}]{chaikin1995principles}%
  \BibitemOpen
  \bibfield  {author} {\bibinfo {author} {\bibfnamefont {P.~M.}\ \bibnamefont {Chaikin}}, \bibinfo {author} {\bibfnamefont {T.~C.}\ \bibnamefont {Lubensky}},\ and\ \bibinfo {author} {\bibfnamefont {T.~A.}\ \bibnamefont {Witten}},\ }\href@noop {} {\emph {\bibinfo {title} {Principles of condensed matter physics}}},\ Vol.~\bibinfo {volume} {10}\ (\bibinfo  {publisher} {Cambridge University Press, Cambridge},\ \bibinfo {year} {1995})\BibitemShut {NoStop}%
\bibitem [{\citenamefont {Cates}\ and\ \citenamefont {Tjhung}(2018)}]{cates2018theories}%
  \BibitemOpen
  \bibfield  {author} {\bibinfo {author} {\bibfnamefont {M.~E.}\ \bibnamefont {Cates}}\ and\ \bibinfo {author} {\bibfnamefont {E.}~\bibnamefont {Tjhung}},\ }\bibfield  {title} {\bibinfo {title} {Theories of binary fluid mixtures: from phase-separation kinetics to active emulsions},\ }\href@noop {} {\bibfield  {journal} {\bibinfo  {journal} {Journal of Fluid Mechanics}\ }\textbf {\bibinfo {volume} {836}},\ \bibinfo {pages} {P1} (\bibinfo {year} {2018})}\BibitemShut {NoStop}%
\bibitem [{\citenamefont {Cates}\ and\ \citenamefont {Tailleur}(2015)}]{cates2015motility}%
  \BibitemOpen
  \bibfield  {author} {\bibinfo {author} {\bibfnamefont {M.~E.}\ \bibnamefont {Cates}}\ and\ \bibinfo {author} {\bibfnamefont {J.}~\bibnamefont {Tailleur}},\ }\bibfield  {title} {\bibinfo {title} {Motility-induced phase separation},\ }\href@noop {} {\bibfield  {journal} {\bibinfo  {journal} {Annu. Rev. Condens. Matter Phys.}\ }\textbf {\bibinfo {volume} {6}},\ \bibinfo {pages} {219} (\bibinfo {year} {2015})}\BibitemShut {NoStop}%
\bibitem [{\citenamefont {Mao}\ \emph {et~al.}(2020)\citenamefont {Mao}, \citenamefont {Chakraverti-Wuerthwein}, \citenamefont {Gaudio},\ and\ \citenamefont {Ko{\v{s}}mrlj}}]{mao2020designing}%
  \BibitemOpen
  \bibfield  {author} {\bibinfo {author} {\bibfnamefont {S.}~\bibnamefont {Mao}}, \bibinfo {author} {\bibfnamefont {M.~S.}\ \bibnamefont {Chakraverti-Wuerthwein}}, \bibinfo {author} {\bibfnamefont {H.}~\bibnamefont {Gaudio}},\ and\ \bibinfo {author} {\bibfnamefont {A.}~\bibnamefont {Ko{\v{s}}mrlj}},\ }\bibfield  {title} {\bibinfo {title} {Designing the morphology of separated phases in multicomponent liquid mixtures},\ }\href@noop {} {\bibfield  {journal} {\bibinfo  {journal} {Physical Review Letters}\ }\textbf {\bibinfo {volume} {125}},\ \bibinfo {pages} {218003} (\bibinfo {year} {2020})}\BibitemShut {NoStop}%
\bibitem [{\citenamefont {Shek}\ and\ \citenamefont {Kusumaatmaja}(2022)}]{shek2022spontaneous}%
  \BibitemOpen
  \bibfield  {author} {\bibinfo {author} {\bibfnamefont {A.~C.}\ \bibnamefont {Shek}}\ and\ \bibinfo {author} {\bibfnamefont {H.}~\bibnamefont {Kusumaatmaja}},\ }\bibfield  {title} {\bibinfo {title} {Spontaneous phase separation of ternary fluid mixtures},\ }\href@noop {} {\bibfield  {journal} {\bibinfo  {journal} {Soft Matter}\ }\textbf {\bibinfo {volume} {18}},\ \bibinfo {pages} {5807} (\bibinfo {year} {2022})}\BibitemShut {NoStop}%
\bibitem [{\citenamefont {Ma}\ and\ \citenamefont {Torquato}(2017)}]{ma2017random}%
  \BibitemOpen
  \bibfield  {author} {\bibinfo {author} {\bibfnamefont {Z.}~\bibnamefont {Ma}}\ and\ \bibinfo {author} {\bibfnamefont {S.}~\bibnamefont {Torquato}},\ }\bibfield  {title} {\bibinfo {title} {Random scalar fields and hyperuniformity},\ }\href@noop {} {\bibfield  {journal} {\bibinfo  {journal} {Journal of Applied Physics}\ }\textbf {\bibinfo {volume} {121}} (\bibinfo {year} {2017})}\BibitemShut {NoStop}%
\bibitem [{\citenamefont {Torquato}\ and\ \citenamefont {Stillinger}(2003)}]{Torquato2003}%
  \BibitemOpen
  \bibfield  {author} {\bibinfo {author} {\bibfnamefont {S.}~\bibnamefont {Torquato}}\ and\ \bibinfo {author} {\bibfnamefont {F.~H.}\ \bibnamefont {Stillinger}},\ }\bibfield  {title} {\bibinfo {title} {Local density fluctuations, hyperuniformity, and order metrics},\ }\href {https://doi.org/10.1103/PhysRevE.68.041113} {\bibfield  {journal} {\bibinfo  {journal} {Phys. Rev. E}\ }\textbf {\bibinfo {volume} {68}},\ \bibinfo {pages} {041113} (\bibinfo {year} {2003})}\BibitemShut {NoStop}%
\bibitem [{\citenamefont {Torquato}(2018)}]{Tor18}%
  \BibitemOpen
  \bibfield  {author} {\bibinfo {author} {\bibfnamefont {S.}~\bibnamefont {Torquato}},\ }\bibfield  {title} {\bibinfo {title} {{Hyperuniform states of matter}},\ }\href {https://doi.org/10.1016/j.physrep.2018.03.001} {\bibfield  {journal} {\bibinfo  {journal} {Phys. Reports}\ }\textbf {\bibinfo {volume} {745}},\ \bibinfo {pages} {1} (\bibinfo {year} {2018})}\BibitemShut {NoStop}%
\bibitem [{\citenamefont {De~Luca}\ \emph {et~al.}(2024)\citenamefont {De~Luca}, \citenamefont {Ma}, \citenamefont {Nardini},\ and\ \citenamefont {Cates}}]{de2024hyperuniformity}%
  \BibitemOpen
  \bibfield  {author} {\bibinfo {author} {\bibfnamefont {F.}~\bibnamefont {De~Luca}}, \bibinfo {author} {\bibfnamefont {X.}~\bibnamefont {Ma}}, \bibinfo {author} {\bibfnamefont {C.}~\bibnamefont {Nardini}},\ and\ \bibinfo {author} {\bibfnamefont {M.~E.}\ \bibnamefont {Cates}},\ }\bibfield  {title} {\bibinfo {title} {Hyperuniformity in phase ordering: the roles of activity, noise, and non-constant mobility},\ }\href@noop {} {\bibfield  {journal} {\bibinfo  {journal} {Journal of Physics: Condensed Matter}\ }\textbf {\bibinfo {volume} {36}},\ \bibinfo {pages} {405101} (\bibinfo {year} {2024})}\BibitemShut {NoStop}%
\bibitem [{\citenamefont {Padhan}\ and\ \citenamefont {Voigt}(2025)}]{padhan2025suppression}%
  \BibitemOpen
  \bibfield  {author} {\bibinfo {author} {\bibfnamefont {N.~B.}\ \bibnamefont {Padhan}}\ and\ \bibinfo {author} {\bibfnamefont {A.}~\bibnamefont {Voigt}},\ }\bibfield  {title} {\bibinfo {title} {Suppression of hyperuniformity in hydrodynamic scalar active field theories},\ }\href@noop {} {\bibfield  {journal} {\bibinfo  {journal} {Journal of Physics: Condensed Matter}\ }\textbf {\bibinfo {volume} {37}},\ \bibinfo {pages} {105101} (\bibinfo {year} {2025})}\BibitemShut {NoStop}%
\bibitem [{\citenamefont {Boyer}\ and\ \citenamefont {Minjeaud}(2014)}]{boyer2014hierarchy}%
  \BibitemOpen
  \bibfield  {author} {\bibinfo {author} {\bibfnamefont {F.}~\bibnamefont {Boyer}}\ and\ \bibinfo {author} {\bibfnamefont {S.}~\bibnamefont {Minjeaud}},\ }\bibfield  {title} {\bibinfo {title} {Hierarchy of consistent n-component {Cahn--Hilliard} systems},\ }\href@noop {} {\bibfield  {journal} {\bibinfo  {journal} {Mathematical Models and Methods in Applied Sciences}\ }\textbf {\bibinfo {volume} {24}},\ \bibinfo {pages} {2885} (\bibinfo {year} {2014})}\BibitemShut {NoStop}%
\bibitem [{\citenamefont {Dong}(2014)}]{dong2014efficient}%
  \BibitemOpen
  \bibfield  {author} {\bibinfo {author} {\bibfnamefont {S.}~\bibnamefont {Dong}},\ }\bibfield  {title} {\bibinfo {title} {An efficient algorithm for incompressible n-phase flows},\ }\href@noop {} {\bibfield  {journal} {\bibinfo  {journal} {Journal of Computational Physics}\ }\textbf {\bibinfo {volume} {276}},\ \bibinfo {pages} {691} (\bibinfo {year} {2014})}\BibitemShut {NoStop}%
\bibitem [{\citenamefont {N{\"u}rnberg}\ \emph {et~al.}(2017)\citenamefont {N{\"u}rnberg} \emph {et~al.}}]{nurnberg2017numerical}%
  \BibitemOpen
  \bibfield  {author} {\bibinfo {author} {\bibfnamefont {R.}~\bibnamefont {N{\"u}rnberg}} \emph {et~al.},\ }\bibfield  {title} {\bibinfo {title} {Numerical approximation of a non-smooth phase-field model for multicomponent incompressible flow},\ }\href@noop {} {\bibfield  {journal} {\bibinfo  {journal} {ESAIM: Mathematical Modelling and Numerical Analysis}\ }\textbf {\bibinfo {volume} {51}},\ \bibinfo {pages} {1089} (\bibinfo {year} {2017})}\BibitemShut {NoStop}%
\bibitem [{\citenamefont {Dong}(2018)}]{dong2018multiphase}%
  \BibitemOpen
  \bibfield  {author} {\bibinfo {author} {\bibfnamefont {S.}~\bibnamefont {Dong}},\ }\bibfield  {title} {\bibinfo {title} {Multiphase flows of n immiscible incompressible fluids: a reduction-consistent and thermodynamically-consistent formulation and associated algorithm},\ }\href@noop {} {\bibfield  {journal} {\bibinfo  {journal} {Journal of Computational Physics}\ }\textbf {\bibinfo {volume} {361}},\ \bibinfo {pages} {1} (\bibinfo {year} {2018})}\BibitemShut {NoStop}%
\bibitem [{\citenamefont {Dong}(2017)}]{dong2017wall}%
  \BibitemOpen
  \bibfield  {author} {\bibinfo {author} {\bibfnamefont {S.}~\bibnamefont {Dong}},\ }\bibfield  {title} {\bibinfo {title} {Wall-bounded multiphase flows of n immiscible incompressible fluids: Consistency and contact-angle boundary condition},\ }\href@noop {} {\bibfield  {journal} {\bibinfo  {journal} {Journal of Computational Physics}\ }\textbf {\bibinfo {volume} {338}},\ \bibinfo {pages} {21} (\bibinfo {year} {2017})}\BibitemShut {NoStop}%
\bibitem [{\citenamefont {Mao}\ \emph {et~al.}(2019)\citenamefont {Mao}, \citenamefont {Kuldinow}, \citenamefont {Haataja},\ and\ \citenamefont {Ko{\v{s}}mrlj}}]{mao2019phase}%
  \BibitemOpen
  \bibfield  {author} {\bibinfo {author} {\bibfnamefont {S.}~\bibnamefont {Mao}}, \bibinfo {author} {\bibfnamefont {D.}~\bibnamefont {Kuldinow}}, \bibinfo {author} {\bibfnamefont {M.~P.}\ \bibnamefont {Haataja}},\ and\ \bibinfo {author} {\bibfnamefont {A.}~\bibnamefont {Ko{\v{s}}mrlj}},\ }\bibfield  {title} {\bibinfo {title} {Phase behavior and morphology of multicomponent liquid mixtures},\ }\href@noop {} {\bibfield  {journal} {\bibinfo  {journal} {Soft Matter}\ }\textbf {\bibinfo {volume} {15}},\ \bibinfo {pages} {1297} (\bibinfo {year} {2019})}\BibitemShut {NoStop}%
\bibitem [{\citenamefont {Abels}\ \emph {et~al.}(2024)\citenamefont {Abels}, \citenamefont {Garcke},\ and\ \citenamefont {Poiatti}}]{abels2024mathematical}%
  \BibitemOpen
  \bibfield  {author} {\bibinfo {author} {\bibfnamefont {H.}~\bibnamefont {Abels}}, \bibinfo {author} {\bibfnamefont {H.}~\bibnamefont {Garcke}},\ and\ \bibinfo {author} {\bibfnamefont {A.}~\bibnamefont {Poiatti}},\ }\bibfield  {title} {\bibinfo {title} {Mathematical analysis of a diffuse interface model for multi-phase flows of incompressible viscous fluids with different densities},\ }\href@noop {} {\bibfield  {journal} {\bibinfo  {journal} {Journal of Mathematical Fluid Mechanics}\ }\textbf {\bibinfo {volume} {26}},\ \bibinfo {pages} {29} (\bibinfo {year} {2024})}\BibitemShut {NoStop}%
\bibitem [{\citenamefont {Kim}\ and\ \citenamefont {Lowengrub}(2005)}]{kim2005phase}%
  \BibitemOpen
  \bibfield  {author} {\bibinfo {author} {\bibfnamefont {J.}~\bibnamefont {Kim}}\ and\ \bibinfo {author} {\bibfnamefont {J.}~\bibnamefont {Lowengrub}},\ }\bibfield  {title} {\bibinfo {title} {Phase field modeling and simulation of three-phase flows},\ }\href@noop {} {\bibfield  {journal} {\bibinfo  {journal} {Interfaces and Free Boundaries}\ }\textbf {\bibinfo {volume} {7}},\ \bibinfo {pages} {435} (\bibinfo {year} {2005})}\BibitemShut {NoStop}%
\bibitem [{\citenamefont {Kim}(2012)}]{kim2012phase}%
  \BibitemOpen
  \bibfield  {author} {\bibinfo {author} {\bibfnamefont {J.}~\bibnamefont {Kim}},\ }\bibfield  {title} {\bibinfo {title} {Phase-field models for multi-component fluid flows},\ }\href@noop {} {\bibfield  {journal} {\bibinfo  {journal} {Communications in Computational Physics}\ }\textbf {\bibinfo {volume} {12}},\ \bibinfo {pages} {613} (\bibinfo {year} {2012})}\BibitemShut {NoStop}%
\bibitem [{\citenamefont {Boyer}\ \emph {et~al.}(2010)\citenamefont {Boyer}, \citenamefont {Lapuerta}, \citenamefont {Minjeaud}, \citenamefont {Piar},\ and\ \citenamefont {Quintard}}]{boyer2010cahn}%
  \BibitemOpen
  \bibfield  {author} {\bibinfo {author} {\bibfnamefont {F.}~\bibnamefont {Boyer}}, \bibinfo {author} {\bibfnamefont {C.}~\bibnamefont {Lapuerta}}, \bibinfo {author} {\bibfnamefont {S.}~\bibnamefont {Minjeaud}}, \bibinfo {author} {\bibfnamefont {B.}~\bibnamefont {Piar}},\ and\ \bibinfo {author} {\bibfnamefont {M.}~\bibnamefont {Quintard}},\ }\bibfield  {title} {\bibinfo {title} {{Cahn--Hilliard/Navier--Stokes} model for the simulation of three-phase flows},\ }\href@noop {} {\bibfield  {journal} {\bibinfo  {journal} {Transport in Porous Media}\ }\textbf {\bibinfo {volume} {82}},\ \bibinfo {pages} {463} (\bibinfo {year} {2010})}\BibitemShut {NoStop}%
\bibitem [{\citenamefont {Boyer}\ and\ \citenamefont {Lapuerta}(2006)}]{boyer2006study}%
  \BibitemOpen
  \bibfield  {author} {\bibinfo {author} {\bibfnamefont {F.}~\bibnamefont {Boyer}}\ and\ \bibinfo {author} {\bibfnamefont {C.}~\bibnamefont {Lapuerta}},\ }\bibfield  {title} {\bibinfo {title} {Study of a three component {Cahn-Hilliard} flow model},\ }\href@noop {} {\bibfield  {journal} {\bibinfo  {journal} {ESAIM: Mathematical Modelling and Numerical Analysis}\ }\textbf {\bibinfo {volume} {40}},\ \bibinfo {pages} {653} (\bibinfo {year} {2006})}\BibitemShut {NoStop}%
\bibitem [{\citenamefont {Hester}\ \emph {et~al.}(2023)\citenamefont {Hester}, \citenamefont {Carney}, \citenamefont {Shah}, \citenamefont {Arnheim}, \citenamefont {Patel}, \citenamefont {Di~Carlo},\ and\ \citenamefont {Bertozzi}}]{hester2023fluid}%
  \BibitemOpen
  \bibfield  {author} {\bibinfo {author} {\bibfnamefont {E.~W.}\ \bibnamefont {Hester}}, \bibinfo {author} {\bibfnamefont {S.}~\bibnamefont {Carney}}, \bibinfo {author} {\bibfnamefont {V.}~\bibnamefont {Shah}}, \bibinfo {author} {\bibfnamefont {A.}~\bibnamefont {Arnheim}}, \bibinfo {author} {\bibfnamefont {B.}~\bibnamefont {Patel}}, \bibinfo {author} {\bibfnamefont {D.}~\bibnamefont {Di~Carlo}},\ and\ \bibinfo {author} {\bibfnamefont {A.~L.}\ \bibnamefont {Bertozzi}},\ }\bibfield  {title} {\bibinfo {title} {Fluid dynamics alters liquid--liquid phase separation in confined aqueous two-phase systems},\ }\href@noop {} {\bibfield  {journal} {\bibinfo  {journal} {Proceedings of the National Academy of Sciences}\ }\textbf {\bibinfo {volume} {120}},\ \bibinfo {pages} {e2306467120} (\bibinfo {year} {2023})}\BibitemShut {NoStop}%
\bibitem [{\citenamefont {Liang}\ \emph {et~al.}(2016)\citenamefont {Liang}, \citenamefont {Shi},\ and\ \citenamefont {Chai}}]{liang2016lattice}%
  \BibitemOpen
  \bibfield  {author} {\bibinfo {author} {\bibfnamefont {H.}~\bibnamefont {Liang}}, \bibinfo {author} {\bibfnamefont {B.}~\bibnamefont {Shi}},\ and\ \bibinfo {author} {\bibfnamefont {Z.}~\bibnamefont {Chai}},\ }\bibfield  {title} {\bibinfo {title} {Lattice {B}oltzmann modeling of three-phase incompressible flows},\ }\href@noop {} {\bibfield  {journal} {\bibinfo  {journal} {Physical Review E}\ }\textbf {\bibinfo {volume} {93}},\ \bibinfo {pages} {013308} (\bibinfo {year} {2016})}\BibitemShut {NoStop}%
\bibitem [{\citenamefont {Abadi}\ \emph {et~al.}(2018)\citenamefont {Abadi}, \citenamefont {Rahimian},\ and\ \citenamefont {Fakhari}}]{abadi2018conservative}%
  \BibitemOpen
  \bibfield  {author} {\bibinfo {author} {\bibfnamefont {R.~H.~H.}\ \bibnamefont {Abadi}}, \bibinfo {author} {\bibfnamefont {M.~H.}\ \bibnamefont {Rahimian}},\ and\ \bibinfo {author} {\bibfnamefont {A.}~\bibnamefont {Fakhari}},\ }\bibfield  {title} {\bibinfo {title} {Conservative phase-field lattice-{B}oltzmann model for ternary fluids},\ }\href@noop {} {\bibfield  {journal} {\bibinfo  {journal} {Journal of Computational Physics}\ }\textbf {\bibinfo {volume} {374}},\ \bibinfo {pages} {668} (\bibinfo {year} {2018})}\BibitemShut {NoStop}%
\bibitem [{\citenamefont {Yang}(2021)}]{yang2021new}%
  \BibitemOpen
  \bibfield  {author} {\bibinfo {author} {\bibfnamefont {X.}~\bibnamefont {Yang}},\ }\bibfield  {title} {\bibinfo {title} {A new efficient fully-decoupled and second-order time-accurate scheme for cahn--hilliard phase-field model of three-phase incompressible flow},\ }\href@noop {} {\bibfield  {journal} {\bibinfo  {journal} {Computer Methods in Applied Mechanics and Engineering}\ }\textbf {\bibinfo {volume} {376}},\ \bibinfo {pages} {113589} (\bibinfo {year} {2021})}\BibitemShut {NoStop}%
\bibitem [{\citenamefont {Kim}(2007)}]{kim2007phase}%
  \BibitemOpen
  \bibfield  {author} {\bibinfo {author} {\bibfnamefont {J.}~\bibnamefont {Kim}},\ }\bibfield  {title} {\bibinfo {title} {Phase field computations for ternary fluid flows},\ }\href@noop {} {\bibfield  {journal} {\bibinfo  {journal} {Computer Methods in Applied Mechanics and Engineering}\ }\textbf {\bibinfo {volume} {196}},\ \bibinfo {pages} {4779} (\bibinfo {year} {2007})}\BibitemShut {NoStop}%
\bibitem [{\citenamefont {Semprebon}\ \emph {et~al.}(2016)\citenamefont {Semprebon}, \citenamefont {Kr{\"u}ger},\ and\ \citenamefont {Kusumaatmaja}}]{semprebon2016ternary}%
  \BibitemOpen
  \bibfield  {author} {\bibinfo {author} {\bibfnamefont {C.}~\bibnamefont {Semprebon}}, \bibinfo {author} {\bibfnamefont {T.}~\bibnamefont {Kr{\"u}ger}},\ and\ \bibinfo {author} {\bibfnamefont {H.}~\bibnamefont {Kusumaatmaja}},\ }\bibfield  {title} {\bibinfo {title} {Ternary free-energy lattice boltzmann model with tunable surface tensions and contact angles},\ }\href@noop {} {\bibfield  {journal} {\bibinfo  {journal} {Physical Review E}\ }\textbf {\bibinfo {volume} {93}},\ \bibinfo {pages} {033305} (\bibinfo {year} {2016})}\BibitemShut {NoStop}%
\bibitem [{\citenamefont {Padhan}\ and\ \citenamefont {Pandit}(2025)}]{padhan2025cahn}%
  \BibitemOpen
  \bibfield  {author} {\bibinfo {author} {\bibfnamefont {N.~B.}\ \bibnamefont {Padhan}}\ and\ \bibinfo {author} {\bibfnamefont {R.}~\bibnamefont {Pandit}},\ }\bibfield  {title} {\bibinfo {title} {The cahn--hilliard--navier--stokes framework for multiphase fluid flows: laminar, turbulent and active},\ }\href@noop {} {\bibfield  {journal} {\bibinfo  {journal} {Journal of Fluid Mechanics}\ }\textbf {\bibinfo {volume} {1010}},\ \bibinfo {pages} {P1} (\bibinfo {year} {2025})}\BibitemShut {NoStop}%
\bibitem [{\citenamefont {W{\"o}hrwag}\ \emph {et~al.}(2018)\citenamefont {W{\"o}hrwag}, \citenamefont {Semprebon}, \citenamefont {Mazloomi~Moqaddam}, \citenamefont {Karlin},\ and\ \citenamefont {Kusumaatmaja}}]{wohrwag2018ternary}%
  \BibitemOpen
  \bibfield  {author} {\bibinfo {author} {\bibfnamefont {M.}~\bibnamefont {W{\"o}hrwag}}, \bibinfo {author} {\bibfnamefont {C.}~\bibnamefont {Semprebon}}, \bibinfo {author} {\bibfnamefont {A.}~\bibnamefont {Mazloomi~Moqaddam}}, \bibinfo {author} {\bibfnamefont {I.}~\bibnamefont {Karlin}},\ and\ \bibinfo {author} {\bibfnamefont {H.}~\bibnamefont {Kusumaatmaja}},\ }\bibfield  {title} {\bibinfo {title} {Ternary free-energy entropic lattice boltzmann model with a high density ratio},\ }\href@noop {} {\bibfield  {journal} {\bibinfo  {journal} {Physical Review Letters}\ }\textbf {\bibinfo {volume} {120}},\ \bibinfo {pages} {234501} (\bibinfo {year} {2018})}\BibitemShut {NoStop}%
\bibitem [{\citenamefont {Park}\ and\ \citenamefont {Anderson}(2012)}]{park2012ternary}%
  \BibitemOpen
  \bibfield  {author} {\bibinfo {author} {\bibfnamefont {J.~M.}\ \bibnamefont {Park}}\ and\ \bibinfo {author} {\bibfnamefont {P.~D.}\ \bibnamefont {Anderson}},\ }\bibfield  {title} {\bibinfo {title} {A ternary model for double-emulsion formation in a capillary microfluidic device},\ }\href@noop {} {\bibfield  {journal} {\bibinfo  {journal} {Lab on a Chip}\ }\textbf {\bibinfo {volume} {12}},\ \bibinfo {pages} {2672} (\bibinfo {year} {2012})}\BibitemShut {NoStop}%
\bibitem [{\citenamefont {Shen}\ and\ \citenamefont {Li}(2023)}]{shen2023bubble}%
  \BibitemOpen
  \bibfield  {author} {\bibinfo {author} {\bibfnamefont {M.}~\bibnamefont {Shen}}\ and\ \bibinfo {author} {\bibfnamefont {B.~Q.}\ \bibnamefont {Li}},\ }\bibfield  {title} {\bibinfo {title} {Bubble rising and interaction in ternary fluid flow: a phase field study},\ }\href@noop {} {\bibfield  {journal} {\bibinfo  {journal} {RSC Advances}\ }\textbf {\bibinfo {volume} {13}},\ \bibinfo {pages} {3561} (\bibinfo {year} {2023})}\BibitemShut {NoStop}%
\bibitem [{\citenamefont {Park}\ and\ \citenamefont {Anderson}(2016)}]{park2016diffuse}%
  \BibitemOpen
  \bibfield  {author} {\bibinfo {author} {\bibfnamefont {J.~M.}\ \bibnamefont {Park}}\ and\ \bibinfo {author} {\bibfnamefont {P.~D.}\ \bibnamefont {Anderson}},\ }\bibfield  {title} {\bibinfo {title} {Diffuse-interface modeling of three-phase interactions},\ }\href@noop {} {\bibfield  {journal} {\bibinfo  {journal} {Applied Physics Letters}\ }\textbf {\bibinfo {volume} {108}} (\bibinfo {year} {2016})}\BibitemShut {NoStop}%
\bibitem [{\citenamefont {Qu{\'e}r{\'e}}\ \emph {et~al.}(1990)\citenamefont {Qu{\'e}r{\'e}}, \citenamefont {Di~Meglio},\ and\ \citenamefont {Brochard-Wyart}}]{quere1990spreading}%
  \BibitemOpen
  \bibfield  {author} {\bibinfo {author} {\bibfnamefont {D.}~\bibnamefont {Qu{\'e}r{\'e}}}, \bibinfo {author} {\bibfnamefont {J.-M.}\ \bibnamefont {Di~Meglio}},\ and\ \bibinfo {author} {\bibfnamefont {F.}~\bibnamefont {Brochard-Wyart}},\ }\bibfield  {title} {\bibinfo {title} {Spreading of liquids on highly curved surfaces},\ }\href@noop {} {\bibfield  {journal} {\bibinfo  {journal} {Science}\ }\textbf {\bibinfo {volume} {249}},\ \bibinfo {pages} {1256} (\bibinfo {year} {1990})}\BibitemShut {NoStop}%
\bibitem [{\citenamefont {Leermakers}\ and\ \citenamefont {Egorov}(2025)}]{leermakers2025no}%
  \BibitemOpen
  \bibfield  {author} {\bibinfo {author} {\bibfnamefont {F.}~\bibnamefont {Leermakers}}\ and\ \bibinfo {author} {\bibfnamefont {S.}~\bibnamefont {Egorov}},\ }\bibfield  {title} {\bibinfo {title} {No violations of critical-point wetting in ternary three fluid-phase systems with short range interactions},\ }\href@noop {} {\bibfield  {journal} {\bibinfo  {journal} {Journal of Colloid and Interface Science}\ }\textbf {\bibinfo {volume} {679}},\ \bibinfo {pages} {124} (\bibinfo {year} {2025})}\BibitemShut {NoStop}%
\bibitem [{\citenamefont {Kovalchuk}\ \emph {et~al.}(2024)\citenamefont {Kovalchuk}, \citenamefont {Sagisaka}, \citenamefont {Komiyama},\ and\ \citenamefont {Simmons}}]{kovalchuk2024spreading}%
  \BibitemOpen
  \bibfield  {author} {\bibinfo {author} {\bibfnamefont {N.~M.}\ \bibnamefont {Kovalchuk}}, \bibinfo {author} {\bibfnamefont {M.}~\bibnamefont {Sagisaka}}, \bibinfo {author} {\bibfnamefont {H.}~\bibnamefont {Komiyama}},\ and\ \bibinfo {author} {\bibfnamefont {M.~J.}\ \bibnamefont {Simmons}},\ }\bibfield  {title} {\bibinfo {title} {Spreading of aqueous surfactant solutions on oil substrates: Superspreaders vs non-superspreaders},\ }\href@noop {} {\bibfield  {journal} {\bibinfo  {journal} {Journal of Colloid and Interface Science}\ }\textbf {\bibinfo {volume} {661}},\ \bibinfo {pages} {1046} (\bibinfo {year} {2024})}\BibitemShut {NoStop}%
\bibitem [{\citenamefont {Cox}\ and\ \citenamefont {Matthews}(2002)}]{cox2002exponential}%
  \BibitemOpen
  \bibfield  {author} {\bibinfo {author} {\bibfnamefont {S.~M.}\ \bibnamefont {Cox}}\ and\ \bibinfo {author} {\bibfnamefont {P.~C.}\ \bibnamefont {Matthews}},\ }\bibfield  {title} {\bibinfo {title} {Exponential time differencing for stiff systems},\ }\href@noop {} {\bibfield  {journal} {\bibinfo  {journal} {Journal of Computational Physics}\ }\textbf {\bibinfo {volume} {176}},\ \bibinfo {pages} {430} (\bibinfo {year} {2002})}\BibitemShut {NoStop}%
\bibitem [{\citenamefont {Tomita}(1991)}]{Tomita1991}%
  \BibitemOpen
  \bibfield  {author} {\bibinfo {author} {\bibfnamefont {H.}~\bibnamefont {Tomita}},\ }\bibfield  {title} {\bibinfo {title} {{Preservation of isotropy at the mesoscopic stage of phase separation processes}},\ }\href@noop {} {\bibfield  {journal} {\bibinfo  {journal} {Prog. Theor. Phys.}\ }\textbf {\bibinfo {volume} {85}},\ \bibinfo {pages} {47–56} (\bibinfo {year} {1991})}\BibitemShut {NoStop}%
\bibitem [{\citenamefont {Steinberg}(1962{\natexlab{a}})}]{doi:10.1073/pnas.48.9.1577}%
  \BibitemOpen
  \bibfield  {author} {\bibinfo {author} {\bibfnamefont {M.~S.}\ \bibnamefont {Steinberg}},\ }\bibfield  {title} {\bibinfo {title} {On the mechanism of tissue reconstruction by dissociated cells, {I}. {P}opulation kinetics, differential adhesiveness, and the absence of directed migration},\ }\href@noop {} {\bibfield  {journal} {\bibinfo  {journal} {Proceedings of the National Academy of Sciences}\ }\textbf {\bibinfo {volume} {48}},\ \bibinfo {pages} {1577} (\bibinfo {year} {1962}{\natexlab{a}})}\BibitemShut {NoStop}%
\bibitem [{\citenamefont {Steinberg}(1962{\natexlab{b}})}]{doi:10.1126/science.137.3532.762}%
  \BibitemOpen
  \bibfield  {author} {\bibinfo {author} {\bibfnamefont {M.~S.}\ \bibnamefont {Steinberg}},\ }\bibfield  {title} {\bibinfo {title} {Mechanism of tissue reconstruction by dissociated cells, {II}: Time-course of events},\ }\href@noop {} {\bibfield  {journal} {\bibinfo  {journal} {Science}\ }\textbf {\bibinfo {volume} {137}},\ \bibinfo {pages} {762} (\bibinfo {year} {1962}{\natexlab{b}})}\BibitemShut {NoStop}%
\bibitem [{\citenamefont {Steinberg}(1962{\natexlab{c}})}]{doi:10.1073/pnas.48.10.1769}%
  \BibitemOpen
  \bibfield  {author} {\bibinfo {author} {\bibfnamefont {M.~S.}\ \bibnamefont {Steinberg}},\ }\bibfield  {title} {\bibinfo {title} {On the mechanism of tissue reconstruction by dissociated cells, {III}. {F}ree energy relations and the reorganization of fused, heteronomic tissue fragments},\ }\href@noop {} {\bibfield  {journal} {\bibinfo  {journal} {Proceedings of the National Academy of Sciences}\ }\textbf {\bibinfo {volume} {48}},\ \bibinfo {pages} {1769} (\bibinfo {year} {1962}{\natexlab{c}})}\BibitemShut {NoStop}%
\bibitem [{\citenamefont {Falc{\'o}}\ \emph {et~al.}(2025)\citenamefont {Falc{\'o}}, \citenamefont {Baker},\ and\ \citenamefont {Carrillo}}]{falco2025nonlocal}%
  \BibitemOpen
  \bibfield  {author} {\bibinfo {author} {\bibfnamefont {C.}~\bibnamefont {Falc{\'o}}}, \bibinfo {author} {\bibfnamefont {R.~E.}\ \bibnamefont {Baker}},\ and\ \bibinfo {author} {\bibfnamefont {J.~A.}\ \bibnamefont {Carrillo}},\ }\bibfield  {title} {\bibinfo {title} {A nonlocal-to-local approach to aggregation-diffusion equations},\ }\href@noop {} {\bibfield  {journal} {\bibinfo  {journal} {SIAM Review}\ }\textbf {\bibinfo {volume} {67}},\ \bibinfo {pages} {353} (\bibinfo {year} {2025})}\BibitemShut {NoStop}%
\end{thebibliography}%
\end{document}